\documentclass[twocolumn]{aastex63}

\graphicspath{{./}{figures/}}

\accepted{May 14, 2021}
\submitjournal{AJ}

\shorttitle{} 
\shortauthors{Wenzl et al.}

\begin{document}

\title{Random Forests as a viable method to select and discover high redshift quasars}

\author[0000-0001-5245-2058]{Lukas Wenzl}
\affiliation{Department of Astronomy, Cornell University, Ithaca, NY, 14853, USA}
\affiliation{Max Planck Institut f\"{u}r Astronomie, K\"{o}nigstuhl 17, D-69117 Heidelberg, Germany}

\author[0000-0002-4544-8242]{Jan-Torge Schindler}
\affiliation{Max Planck Institut f\"{u}r Astronomie, K\"{o}nigstuhl 17, D-69117 Heidelberg, Germany}

\author[0000-0003-3310-0131]{Xiaohui Fan}
\affiliation{Steward Observatory, University of Arizona, 933 N Cherry Ave, Tucson, AZ 85721, USA}

\author[0000-0001-6102-9526]{Irham Taufik Andika}
\affiliation{Max Planck Institut f\"{u}r Astronomie, K\"{o}nigstuhl 17, D-69117 Heidelberg, Germany}
\affiliation{International Max Planck Research School for Astronomy \& Cosmic Physics at the University of Heidelberg}

\author[0000-0002-2931-7824]{Eduardo Ba\~nados}
\affiliation{Max Planck Institut f\"{u}r Astronomie, K\"{o}nigstuhl 17, D-69117 Heidelberg, Germany}

\author[0000-0002-2662-8803]{Roberto Decarli}
\affil{INAF --- Osservatorio di Astrofisica e Scienza dello Spazio, via Gobetti 93/3, I-40129, Bologna, Italy}

\author[0000-0003-3804-2137]{Knud Jahnke}
\affiliation{Max Planck Institut f\"{u}r Astronomie, K\"{o}nigstuhl 17, D-69117 Heidelberg, Germany}

\author[0000-0002-5941-5214]{Chiara Mazzucchelli}\thanks{ESO Fellow}
\affiliation{European Southern Observatory, Alonso de Cordova 3107, Vitacura, Region Metropolitana, Chile}

\author[0000-0003-2984-6803]{Masafusa Onoue}
\affiliation{Max Planck Institut f\"{u}r Astronomie, K\"{o}nigstuhl 17, D-69117 Heidelberg, Germany}

\author[0000-0001-9024-8322]{Bram P.\ Venemans}
\affiliation{Leiden Observatory, Leiden University, PO Box 9513, 2300 RA, Leiden, The Netherlands}

\author[0000-0003-4793-7880]{Fabian Walter}
\affiliation{Max Planck Institut f\"{u}r Astronomie, K\"{o}nigstuhl 17, D-69117 Heidelberg, Germany}

\author[0000-0001-5287-4242]{Jinyi Yang}
\affiliation{Steward Observatory, University of Arizona, 933 N Cherry Ave, Tucson, AZ 85721, USA}





\begin{abstract}
We present a method of selecting quasars up to redshift $\approx$ 6 with random forests, a supervised machine learning method, applied to Pan-STARRS1 and WISE data. We find that, thanks to the increasing set of known quasars we can assemble a training set that enables supervised machine learning algorithms to become a competitive alternative to other methods up to this redshift. 
We present a candidate set for the redshift range 4.8 to 6.3 which includes the region around z = 5.5 where quasars are difficult to select due to photometric similarity to red and brown dwarfs. We demonstrate that under our survey restrictions we can reach a high completeness ($66 \pm 7 \%$ below redshift 5.6 / $83^{+6}_{-9}\%$ above redshift 5.6) while maintaining a high selection efficiency ($78^{+10}_{-8}\%$ / $94^{+5}_{-8}\%$). Our selection efficiency is estimated via a novel method based on the different distributions of quasars and contaminants on the sky. The final catalog of 515 candidates includes 225 known quasars. We predict the candidate catalog to contain additional $148^{+41}_{-33}$ new quasars below redshift 5.6 and $45^{+5}_{-8}$ above and make the catalog publicly available.
%
Spectroscopic follow-up observations of 37 candidates lead us to discover 20 new high redshift quasars (18 at $4.6\le z\le5.5$, 2 $z\sim5.7$). These observations are consistent with our predictions on efficiency. We argue that random forests can lead to higher completeness because our candidate set contains a number of objects that would be rejected by common color cuts, including one of the newly discovered redshift 5.7 quasars. 


\end{abstract}



\section{Introduction}\label{sec:intro}


Large samples of luminous high redshift quasars not only allow us to study the onset of black hole growth and supermassive black hole formation \citep{Volonteri2012}. They are essential probes to study the evolution of the intergalactic medium when the universe was only around a billion years old. For example, measurements of the Gunn-Peterson trough in spectra of quasars at $z\approx 6$ indicate a rapid increase in the fraction of neutral Hydrogen between redshift 5.5 and 6, putting strong constraints on the end of cosmic re-ionization \citep{1965ApJ...142.1633G,Becker2001,Fan2006,McGreer2015}.

The Quasar Luminosity Function (QLF) traces the spatial density of quasars throughout cosmic time and helps us to understand the evolution of supermassive black holes \citep{Schmidt1968,Boyle2000,Croom2004,Ross2013}. At high redshift, small sample sizes lead to large uncertainties in the determination of the exact shape and evolution of the QLF \citep{Jiang2008,Venemans2013,Kashikawa2015,Matsuoka2018LuminosityFunction,Wang2019,Yang2019}. Nonetheless, current results still allow for physical conclusions: for example, quasars are likely not the main producers of re-ionization photons \citep{Willott2010,McGreer2013}. The QLF can also be used to estimate the number of quasars future surveys will be able to find \citep[e.g.][]{Willott2010}. 

All of these studies rely on well-defined spectroscopically confirmed quasar samples. Therefore, we must be able to identify and confirm high redshift quasars with a well-defined selection function and maximize efficiency and completeness to best use limited observational resources. 

To date around $8 \cdot 10^5$ quasars have been spectroscopically identified through a wide range of efforts \citep{Schmidt1963,Hewett1995,Boyle2000,Richards2002,Dawson2013,Dawson2016,Lyke2020}. While most of them are found at low to intermediate redshifts there have been several specialized efforts to find high redshift quasars in large-area surveys. For this work, we define high redshift to mean $z>4.7$. This class of high redshift quasars has now thousands of known objects with contributions among others from \citet{Fan2000}, \citet{McGreer2013}, \cite{Wang2016}, \cite{Banados2016}, \cite{Jiang2016}, \cite{Yang2017}, \cite{Matsuoka2018}, and  \cite{Yang2019}. 


At redshifts above $z=4.7$ the Lyman-$\alpha$ emission line is significantly redshifted into the i band and even redder wavelengths. Furthermore, the blue-ward flux is absorbed by the intervening hydrogen creating the so-called Lyman break in the spectrum.
It is therefore essential to use infrared photometry to constrain the quasar continuum. Combining infrared with optical photometry then enables one to detect the Lyman break differentiating these quasars further from other objects.


Many selection methods for high redshift quasars make use of the broadband colors and magnitudes of large photometric catalogs and combine them with information about the morphology, time variability, X-ray or radio detections, position and proper motion \citep{Banados2016,Palanque-Delabrouille2011,Assef2011,Kozlowski2019,Banados2015Constraining,Banados2018Reionization,McGreer2009,Bailer-Jones2019}. 
Sophisticated color cuts define selection regions in color-color space to separate quasar and contaminant distributions \citep[e.g.][]{Richards2002}. This leads to well-defined selections that are easily reproducible and can be justified with physical reasoning (e.g. the redshift evolution of the Lyman-$\alpha$ emission through the broadband filters). However, color cuts might not make use of all the available information by ignoring correlations in the full high dimensional color space. Furthermore, they represent hard cuts potentially missing quasars scattering out of the selection regions, which could be remedied by a more probabilistic approach \citep[e.g.][]{Mortlock2012}. On the other hand, the majority of high redshift quasars have been found by using color selection criteria \citep{Banados2016, Yang2017}. Often, simulations of high redshift quasars were used to inform these color cuts \citep{McGreer2013}.  

Another method to exploit the photometric information of large surveys is spectral energy distribution (SED) fitting. The best fits to templates of appropriately red-shifted quasar spectra are often compared with best fits of their main contaminants \citep{Reed2017}. This method relies on a correct understanding of the evolution of quasars and also the most common contaminants but makes effective use of the photometric information.

Machine learning methods have been successfully employed to select quasars up to intermediate redshift $z\sim$4.7 \citep{Richards2009,Bovy2011,Jin2019,Khramtsov2019}. 
%
%
From a range of available methods, we adopt random forests, a supervised machine learning approach that has been used for quasar selection successfully \citep{Schindler2017,Nakoneczny2019,2020RNAAS...4..179Y}. We choose random forests for their robustness and fast training but expect that we could achieve comparable results with other common approaches. In recent comparisons for quasar searches, random forest achieved similar results as Support Vector Machines, XGBoost, and Artificial Neural Networks \citep{Schindler2017,2019A&A...632A..56K,2020arXiv201013857N}. Our main focus will be to demonstrate that we can successfully extend a supervised machine learning approach to the high-redshift domain even though the training samples are significantly smaller than at lower redshift. 

In the following, we will demonstrate that there are enough known objects in this class to effectively train a random forest algorithm to select these quasars using photometric data from the
Panoramic Survey Telescope and Rapid Response System \citep[Pan-STARRS;][]{Chambers2016} and the Wide-field Infrared Survey Explorer \citep[WISE; ][]{Wright2010} up to redshifts of 6 while only missing objects in a relatively small range around $z\approx 5.4$. In Section\,\ref{section:data_preparation} we discuss the catalog data we use and how we assemble our training set. In Section\,\ref{section:random_forest} we introduce the random forest selection approach and evaluate it via cross-validation. In Section\,\ref{sec_eff} we discuss a method to predict the efficiency of our selection and in Section\,\ref{section:selection_results} we present the resulting high-$z$ quasar selection. In Section\,\ref{section:observations_results} we present the results of the observation of some of our candidates. These include the discovery of 20 new high-$z$ quasars. We discuss our results and summarize our findings in Section\,\ref{section:conclusion}.

Unless otherwise noted all magnitudes are given in the AB system and are already corrected for galactic extinction using the  \citet{Schlegel1998} dust map with the updated filter corrections from \citet{Schlafly2011}\footnote{The filter corrections for WISE W1 and W2 are extrapolated values taken from IRSA, see for example \href{https://irsa.ipac.caltech.edu/workspace/TMP_toDGRk_31798/DUST/M_31.v0001/extinction.html}{irsa.ipac.caltech.edu/workspace/TMP\_toDGRk\_31798/} }. Furthermore we use a standard flat $\Lambda$CDM cosmology with $\Omega_\Lambda = 0.7$, $\Omega_m = 0.3$ and $H_0 = 70 \text{ km s}^{-1} \text{Mpc}^{-1}$.


\section{Data preparation}\label{section:data_preparation}
\subsection{Catalog data}

The data we are mining for quasars is a cross-match between the publicly available Pan-STARRS DR1 described in \citet{Chambers2016} and ALLWISE \citep{Cutri2014} catalogs. The ALLWISE survey is a release of the aggregated data from WISE and its extended mission NEOWISE \citep{Mainzer2011} up until 2013. From Pan-STARRS we use the five stacked PSF magnitudes ($\text{g}_{\text{PSF}}\text{, r}_{\text{PSF}}\text{, i}_{\text{PSF}}\text{, z}_{\text{PSF}}\, \text{and y}_{\text{PSF}}$), stacked aperture magnitude in the z band ($\text{z}_{\text{APERTURE}}$), mean position and the objectinfoFlag. From WISE we use the 3.4 and 4.6$\mu$m broadband magnitudes ($\text{W}1, \, \text{W}2$), their signal to noise ratio ($\text{W}1_{s/n}, \text{W}2_{s/n} $), position, the active deblending flag (\texttt{na}) and the number of PSF components used for the PSF fitting (\texttt{nb}).

We use the python framework Large Survey Database \citep{Juric2012} to cross-match the two catalogs applying the following selection criteria:
%
\begin{eqnarray}
14 < \text{z}_{\text{PSF}} \leq 20.5 \\\label{cond_ps_start}
\text{y}_{\text{PSF}} \text{ is not None} \\
-0.3 \leq \text{z}_{\text{PSF}} - \text{z}_{\text{APERTURE}}  \leq 0.3 \label{exfilter} \\
\text{galactic latitude}   > 20^\circ \text{ or} <-20^\circ \\\label{cond_ps_end}
\text{objinfoFlag has \texttt{GOOD} and \texttt{GOOD\_STACK}} \\
\text{2\farcs0 match in ALLWISE} \\
\text{W}1_{s/n} \geq 5 \\
\text{W}2_{s/n} \geq 3 \\
\texttt{na} = 0, \texttt{nb} = 1
\end{eqnarray}
%
%
The resulting catalog has around 72 million objects. The z band is used to select the brightness range from 14 to 20.5 in magnitude. Since the brightest quasar at $z\geq4.7$ in our training data (see Section\,\ref{subsection:training_set}) has a z band magnitude of $\text{z}_{\text{PSF}} = 17.3$, there is only a remote chance to miss quasar lenses by adopting a bright limit for our selection. We also choose a faint limiting magnitude on the z band that is well above the detection limit to ensure the reliability of the photometry. The 5 sigma detection limit for the Pan-STARRS survey in the stacked z band for point sources is 22.3 mag in AB \citep[see Table 11,][]{Chambers2016}. They also showed that in the z band the 98\% source completeness limit is fainter than 20.5 mag on most of the sky, especially away from the galactic plane \citep[see Figure 17,][]{Chambers2016}. The criteria on the z band automatically remove all objects with a missing z band detection. We further require the y band not to be "None". However, the other bands of Pan-STARRS (g, r and i) can be missing because we expect the targeted high redshift quasars to have very little flux in these bluer bands. 

We use the difference of the PSF and aperture magnitude in the z band to actively exclude sources with extended morphologies from our selection. The cutoff of $\text{z}_{\text{PSF}} - \text{z}_{\text{APERTURE}} = \pm 0.3$ is informed by Fig.~3 in \cite{Banados2016}, where the magnitude difference is compared for stars, quasars, and galaxies. This cut is designed to effectively remove galaxies from our selection, however, it may also reduce our sensitivity to lensed quasars. In our final candidate selection (see Section~\ref{section:selection_results}) there are only a few remaining galaxies that were removed during visual selection, so this approach is sufficient for our purpose.

We furthermore restrict our selection to Galactic latitudes of $|b|\ge20^\circ$, where the contamination by galactic stars decreases significantly. 
We require the photometry to fulfill the \texttt{GOOD} and \texttt{GOOD\_STACK} flags in the \texttt{objectinfoFlag} from Pan-STARRS. These are quality flags provided by Pan-STARRS to indicate that the object has a good-quality measurement in the data and a good-quality object in the stack ($>1$ good stack measurement). We matched our objects with ALLWISE with a radius of $2\farcs0$ using only the closest match. We require that $\text{W}1$ and $\text{W}2$ are detected with a signal-to-noise of 5 and 3 respectively. We exclude obviously blended sources via the active deblending flag (\texttt{na}) and the number of PSF components used for the PSF fitting (\texttt{nb}) from WISE. These WISE flags ensure more reliable photometry but we note that they reduce our sensitivity to lensed quasars and may remove some quasars with close-by sources.

To determine which survey limits our quasar selection we consider the set of \textit{Additional High Redshift Quasars}, discussed in the next section. Of a total of 1001 quasars, 936 have a Pan-STARRS match, 657 full-fill conditions (1-3) mostly limited by the brightness cut in the z band. 

We contrast this with 710 objects that have an ALLWISE match and full-fill ALLWISE photometry conditions (6,7), 647 additionally full-fill condition (8). Both the Pan-STARRS and ALLWISE photometry requirements remove a similar fraction of known quasars and the remaining objects have a large overlap: 565 objects full-fill conditions (1-3,6-8). This shows that our requirements on both surveys are well balanced for our targeted class. To use fainter objects in the Pan-STARRS data we would also require deeper infrared data. We note that in Table~\ref{datasets} we only list additional quasars that are not already in the other set. 


\subsection{Training data} \label{subsection:training_set}

Random forests are a supervised machine learning algorithm and therefore heavily rely on representative training sets. It is essential to assemble a training set consisting of spectroscopically identified objects representing the wide range of different objects in our catalog data. 
For a reliable selection of high redshift quasars at $z=4.7-6$ we need to make sure to construct a representative training set. This means that we need to include all potential contaminants that populate the same color space, like M-, L-, and T-dwarfs.
We do not include galaxies in our training set since we already removed extended sources from our data set (see selection criterion 3). The training classes used in our random forest selection are listed in Table~\ref{tab:classes}. 
\begin{table}[h]
\centering
\begin{tabular}{ccccccc}
\toprule
------ stars ------ & \multicolumn{4}{c}{-------------------- quasars --------------------}\\
A F G K M L T & vlow-$z$ & low-$z$ & mid-$z$ & high-$z$  \\
 & $(0, 1.5]$ & $(1.5, 3.5]$ & $(3.5,4.7]$ & $z > 4.7$ \\
\toprule
\end{tabular}
\caption{\label{tab:classes}Classes used for the random forest classification: A to T type stars and 4 redshift bins for quasars (redshift ranges given below the classes). The goal is to find objects in the high-$z$ bin.}
\end{table}
%

We do not include O and B type stars as they are far from high redshift quasars in color space. These classes are irrelevant since they likely get assigned the label of the most similar star class, but are not confused with our targeted quasars. We exclude the Y type brown dwarfs since we do not have many objects of that class and they are also not relevant contaminants for quasars with $z\sim 5-6$. Similarly, there are classes of objects that are underrepresented in our training set like low redshift BAL quasars which are known contaminants for high redshift quasars. 


\newcommand{\ra}[1]{\renewcommand{\arraystretch}{#1}}
\begin{table}
\ra{1.3}
\centering
\begin{tabular}{@{}p{25mm}ccccc@{}} 
\toprule 
~ &\multicolumn{5}{c}{\# of stars and quasars}\\\cline{2-6}
Description        & A-K      & M & L,T & $z\leq 4.7$  & $z > 4.7$      \\\toprule
Schindler+2019  & 2.0E5  & 5.8E4 & 1145  & 1.6E5 & 129 \\
Dwarfs    &    -     & 34 & 436 &   -   & -    \\
Additional High \newline Redshift Quasars  & - &- & - & 137 & 337        \\
\toprule
\end{tabular}
\caption{\label{datasets}The data for the training set full-filling our photometric restrictions. The main part is adapted from \citet{Schindler2019} based on SDSS data. The additional red and brown dwarfs are from the Dwarf Archive and the Additional High Redshift Quasars are an assemblage of recent surveys.}
\end{table}

\begin{figure}
\plotone{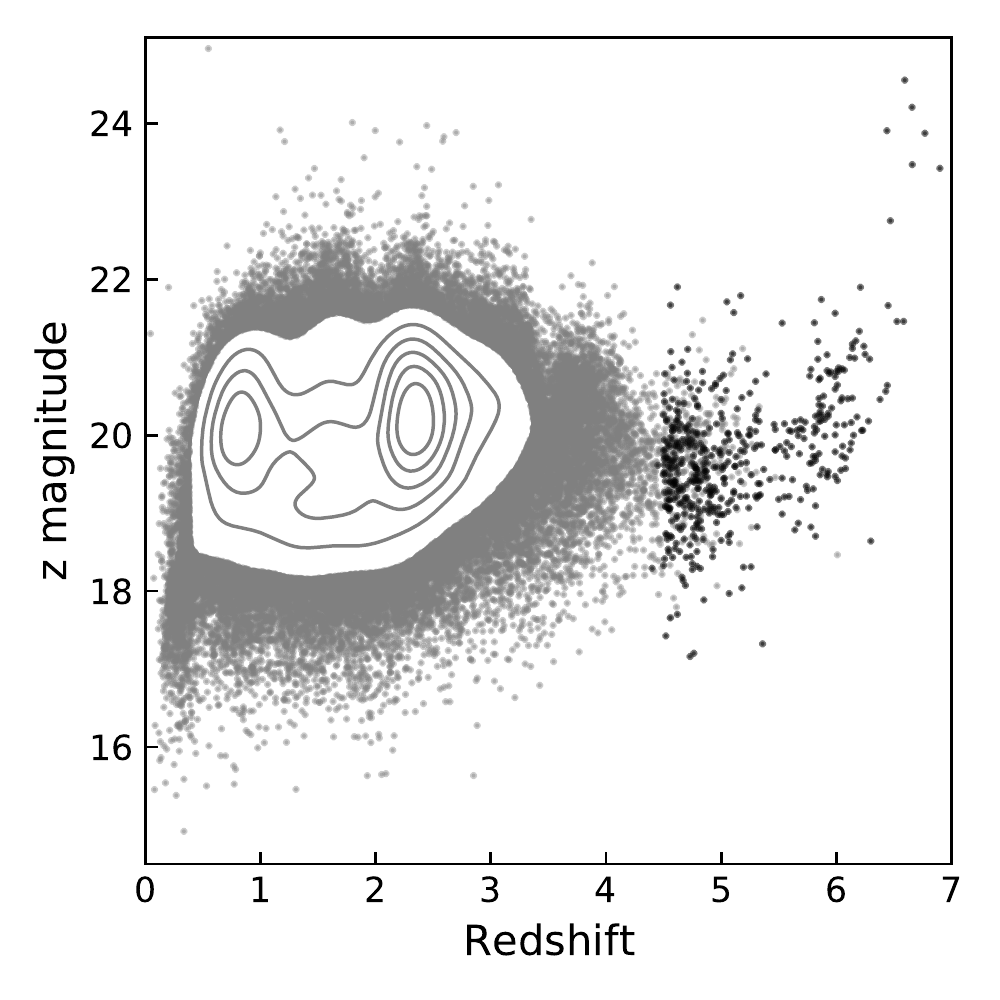}
\caption{Dust corrected Pan-STARRS PSF magnitude in the z band vs redshift for all known quasars in the training data. In grey the \citet{Schindler2019} dataset with the densest part shown as density contours. In black the additional quasars from high redshift surveys. We note that there is an under density of known quasars around redshift 5.5. \label{fig:training_qso}}
\end{figure}

\begin{figure*}
\includegraphics[width= 1. \textwidth]{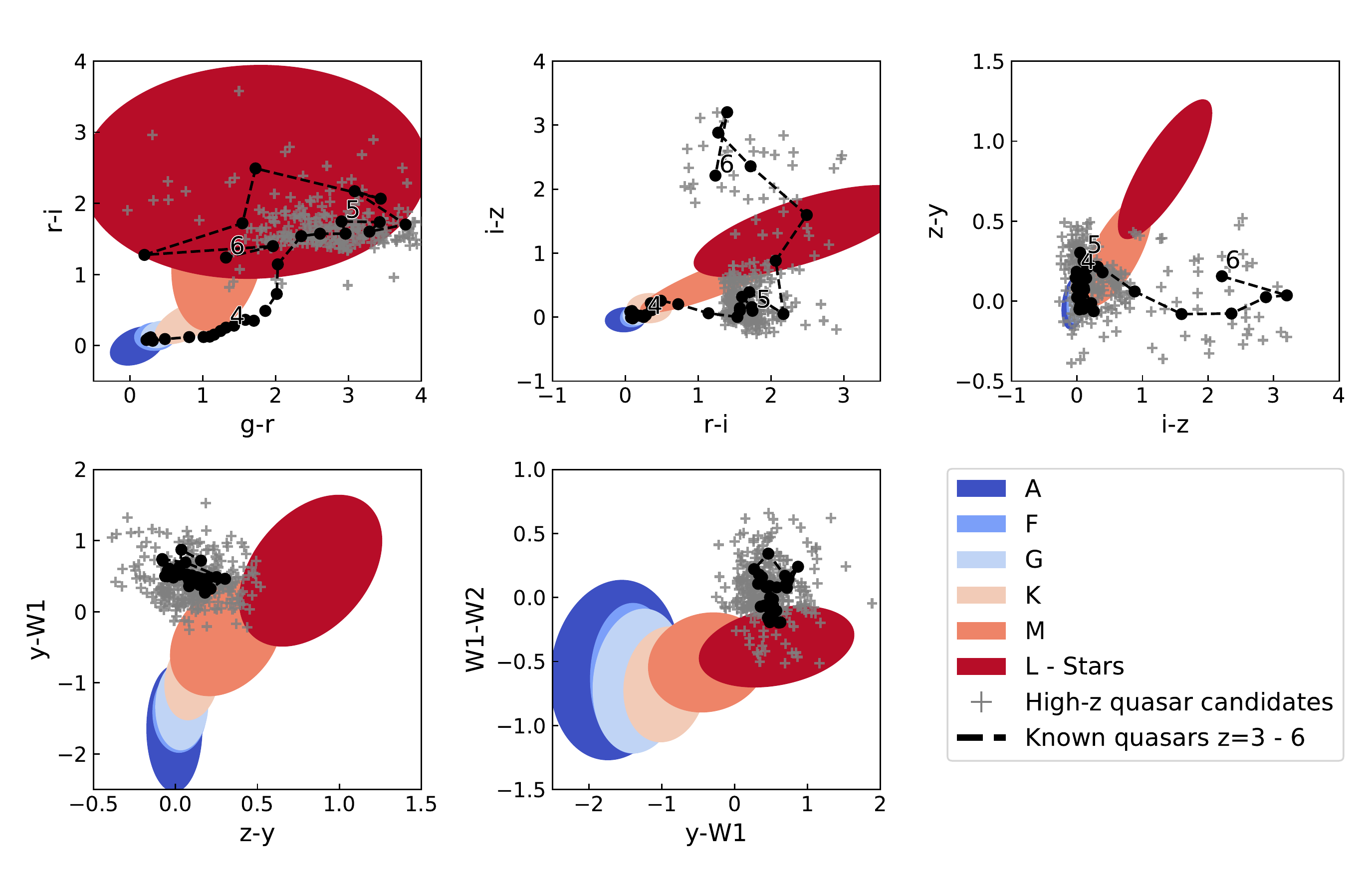}
\caption{Summary plot highlighting the color information we use to separate stars and quasars as well as to estimate the redshift of quasar candidates. We show color-color plots for the Pan-STARRS and WISE bands we use. All magnitudes are in AB and extinction corrected. We show ellipses for each stellar class containing about 95\% of the training set. The T class is not shown as its ellipse would be too large. 
The quasar track in black shows the redshift evolution of quasar in color space. To find the track we averaged quasar colors in our training set binned in steps of $\Delta z=0.1$.
%
Finally, as grey pluses, we show the \textit{high-$z$ candidate set} that we publish with this paper and describe in Section~\ref{section:selection_results}. It contains 515 promising quasar candidates at high redshift. Candidates that are missing detections in the g, r or i bands are not shown.\label{fig:all_color_color_plots}}
\end{figure*}

We built our training set with the spectroscopic training set from \cite{Schindler2019}. It is based on the spectroscopically confirmed quasars from the Sloan Digital Sky Survey (SDSS) DR7 and DR12 quasar catalogs as well as the spectroscopically confirmed stars from SDSS DR 13. The SDSS data was matched to Pan-STARRS within $3\farcs98$, using only the closest match. For a full discussion of the data processing, see the referenced paper. We take only the Pan-STARRS position, classification as well as the redshift for the quasars and reprocess the rest of the data for internal consistency.
We expand this training set with more objects in the relevant color space region. To increase the number of red and brown dwarfs the Dwarf Archive\footnote{\href{http://dwarfarchives.org/}{dwarfarchives.org/}} is used. We match their positions to Pan-STARRS within $2\farcs0$. This should be sufficient for our purposes but we acknowledge that a more careful cross-match considering proper motion would increase the number of dwarfs further. 
To supplement our training set with additional high-$z$ quasars, we added a comprehensive list of quasars known as of mid-2018. Again we cross-match the position to Pan-STARRS within $2\farcs0$. The major sources are \citet{Wang2016}, \citet{Banados2016}, \citet{Jiang2016}, \citet{McGreer2013}, \citet{Matsuoka2018}, \citet{Yang2017}, and the preliminary results of  \citet{Yang2019} as of mid 2018. 

For each of our three datasets, we download the Pan-STARRS data from MAST using the CasJobs\footnote{\href{http://casjobs.sdss.org/CasJobs}{casjobs.sdss.org/CasJobs}} interface. We cross-match with ALLWISE with a radius of $2\farcs0$ using IRSA\footnote{\href{https://irsa.ipac.caltech.edu/applications/Gator/}{irsa.ipac.caltech.edu/applications/Gator/}} and only using the closest match. To include as many training objects as possible, we omit some of our photometric selection criteria used for the full catalog data. The classification information added outweighs the downside of them not being fully representative of the catalog data we collected. We require conditions (2, 6-7) as well as a detection in the Pan-STARRS z and WISE $\text{W}2$ bands which means a value entry in the catalog. Finally, we remove duplicates between our three datasets based on the Pan-STARRS ObjID. 

In total the \cite{Schindler2019} dataset gives us 259,240 stars and 164,318 quasars that full-fill our photometric requirements. 
In addition to the SDSS DR7 and DR12 quasars, we add a total of 474 additional quasars from high redshift surveys. We also add an additional 470 dwarfs from the Dwarf Archive.
Table~\ref{datasets} lists the different sources and classifications for the training set with the number of usable objects they add. In Fig.~\ref{fig:training_qso} we show a plot of the dust corrected z band magnitude from Pan-STARRS vs redshift for all training quasars. Outliers beyond magnitude 25 are cut off. At the high redshift end, the additional quasars from recent surveys significantly extend the training set. We note that there is a visible under-density of known quasars around redshift 5.5. At this redshift quasars are challenging to find as their colors in common bands are very similar to those of M stars \citep{Yang2019}. 

In Fig.~\ref{fig:all_color_color_plots} we show color-color plots for the objects in our training set. This highlights that the Pan-STARRS and WISE bands contain information that will allow us to differentiate the classes listed in Table~\ref{tab:classes}. We note that this visualization emphasizes how the average color information of quasars differs from stars and evolves with redshift, but hides the complexity of applying this to real data. The 5\% of stars with larger scatter than the ellipses shown far outnumber the high-$z$ quasars and the quasars themselves are also scattered. To reliably classify new objects we need the full high dimensional color information.

\subsection{Data pre-processing}
We correct both our catalog data and training set for galactic extinction based on the dust map of \cite{Schlegel1998}, using the sfdmap\footnote{\href{https://github.com/kbarbary/sfdmap}{github.com/kbarbary/sfdmap}} python package. 
The Vega magnitudes in the ALLWISE catalog are converted to AB magnitudes using the constants $\text{W}1_{\textrm{AB}} - \text{W}1_{\textrm{Vega}} = 2.699$ and $ \text{W}2_{\textrm{AB}} - \text{W}2_{\textrm{Vega}} = 3.339$ \citep[Sec IV 4h,][]{2012wise.rept....1C}. All magnitudes are then converted to flux density in Jansky units. 
Our catalog restrictions allow objects with non-detections in the g, r and i bands to be considered. However, the random forest method can not handle null values. We work around this by replacing the missing values with a fixed value that is fainter than the detection limit of the catalog. This way the resulting flux density ratios will be close to the true values. We choose 1e-10 Jy or a magnitude of 33.90 in the AB system as the replacement value. 
We replace all missing g, r and i band measurements with this value. 

For our analysis we consider the flux densities $F_{\text{g}}$, $F_{\text{r}}$, $F_{\text{i}}$, $F_{\text{z}}$, $F_{\text{y}}$, $F_{\text{W}1}$, $F_{\text{W}2}$ and flux density ratios $F_{\text{g}}/ F_{\text{r}}$, $F_{\text{r}}/ F_{\text{i}}$, $F_{\text{i}}/ F_{\text{z}}$, $F_{\text{z}}/ F_{\text{y}}$, $F_{\text{y}}/ F_{\text{W}1}$, $F_{\text{W}1}/ F_{\text{W}2}$ as features. In Section~\ref{sec:featureselection} we choose a subset of these based on their individual information contribution.



\section{Random Forest Selection}
\label{section:random_forest}

In this section, we present our approach of selecting candidates with random forests, a popular method for supervised machine learning \citep{Ho1995,Breiman2001}. We first use a random forest classifier to separate our catalog data into the classes from Table~\ref{tab:classes} and then a random forest regressor to find a redshift estimate for the most promising candidates. We briefly describe how the algorithm works, introduce the common metrics to evaluate the classification/regression, and then discuss our cross-validation results.

\subsection{Random Forests introduction}

The random forest algorithm trains a large set of binary decision trees using a training set with a set of features and known classes or redshift. Each binary tree makes a prediction for the probability distribution of classes or the expected redshift for our quasar candidates. In the sci-kit learn implementation \citep{Pedregosa2011} adopted here, the final pseudo-probability distribution for the classes or the expected redshift is the average of the prediction from each tree. We decide to use the photometric information in the form of flux density ratios as well as the two flux densities, resulting in k=8 features.

The binary decision trees are built from the training set by determining the best cut along one feature axis via a minimization problem. For the classification, we minimize the Gini impurity $\left(G:= 1 - \sum_i^k p_i^2\right)$ and for the regression the sum of squared errors in redshift. This cut will split the sample in the current node of the tree into two subsamples, its children. The remaining objects in each child give the probabilities for each class as their percentage share of the leaf ($p_i$) or the redshift estimate through the average redshift of the objects in the child. The tree will be developed until a stopping condition is reached (e.g. a specified minimum sample size per child node or the maximum depth of the tree). For a quasar candidate, the prediction is based on the $p_i$ or average redshift of the deepest child it belongs to.

Single decision trees are prone to overfit the training data. Hence, a random forest uses ensembles of randomly built decision trees to counteract overfitting. The source of randomization is two-fold: 1) Individual bootstrap samples from the training set are drawn to build the trees. 2) Only a subset ($ \lfloor \sqrt{k} \rfloor$ in our case) of all k features is considered to find the best split for each internal node. This decreases the running time and reduces the correlation between the individual trees further than just training on the bootstrap samples of the training data. Otherwise, features that are strong predictors would be cut very similar for most trees and thereby result in correlated trees. This was empirically demonstrated by \cite{Ho1998}. Correlated trees are undesired since the underlying assumption to be able to average the trees is that they are independent.

One of the main advantages of random forests is that their training is relatively fast. They can run in parallel since the different decision trees can be calculated independently and they scale well. A random forest with T trees and N training objects takes $O(T \, N \, \text{log} \, N)$ time to build and can be applied in $O(T \, \text{log} \, N)$ time.

More details about the random forest algorithm used can be found in \citet[Chapter 14]{Bishop2006} and \citet[Chapter 9]{Ivezic2014MOREINFO}. For this work the implementation of the random forest classifier and regressor in scikit-learn (version 0.19.1) by \citet{Pedregosa2011} for python is used. The hyper-parameters \texttt{min\_sample\_split}, \texttt{max\_depth} and \texttt{n\_estimators} were optimized for the training set using scikit-learn's \texttt{GridSearch} function. All unmentioned other hyper-parameters are left at their default values.



\subsection{Terminology}
\label{sec:definitions}

To evaluate the performance of the classification we use the two measures \textit{recall} and \textit{precision}. To evaluate them we use cross-validation. For this we split our training set into two parts, train the random forest on one part and then predict the classes for the other part. Considering the resulting true positives ($T_p$), false positives ($F_p$) and false negatives ($F_n$) \textit{recall} is defined as
\begin{equation}
R := \frac{T_p}{T_p + F_n}
\end{equation}
and \textit{precision} is defined as
\begin{equation}\label{def_precision}
P := \frac{T_p}{T_p+F_p}.
\end{equation}

We are only interested in objects in the high-$z$ class, so we consider the objects put into this class as the positives and all others as the negatives. 
Each candidate is given a probability to belong to the high-$z$ class by the random forest classification. We set a cut-off probability for the high-$z$ class to decide whether an object will be a valid candidate in our selection. Changing this cutoff probability allows us to increase the \textit{recall}. However, in return, this will lower the \textit{precision} of our classification. This is why we need both parameters to evaluate the performance of the random forest fully.

We interpret the \textit{recall} as an estimate of the completeness of our selection, i.e. the fraction of all high-$z$ quasars inherent in our photometric selection the random forest correctly classifies. Similarly, we interpret the \textit{precision} as an upper limit of the efficiency, where the efficiency is the fraction of the final candidates that are high-$z$ quasars. We will use the terms completeness and upper limit on efficiency in the following and give a justification for our interpretation in Section~\ref{crossvaldiationsection}.

\subsection{Feature selection}\label{sec:featureselection}

\begin{table}[]
\centering
\begin{tabular}{lc}
\toprule
Feature                             & \multicolumn{1}{l}{Importance {[}\%{]}} \\
\toprule
$F_{\text{g}}/ F_{\text{r}}$        & 17                                      \\
$F_{\text{y}}/ F_{\text{W}1}$       & 17                                      \\
$F_{\text{r}}/ F_{\text{i}}$        & 14                                      \\
$F_{\text{i}}/ F_{\text{z}}$        & 9                                       \\
$F_{\text{z}}/ F_{\text{y}}$        & 9                                       \\
$F_{\text{W}1}/ F_{\text{W}2}$      & 8                                       \\
$F_{\text{i}}$ & 4                                       \\
$F_{\text{W}2}$                     & 4                                       \\
$F_{\text{y}}$                      & 4                                       \\
$F_{\text{W}1}$                     & 4                                       \\
$F_{\text{z}}$                      & 4                                       \\
$F_{\text{r}}$                      & 3                                       \\
$F_{\text{g}}$                      & 3 \\
\toprule
\end{tabular}
\caption{\label{table:feature_importance} To select the features we run the random forest classification with all available flux densities and ratios. The importance is calculated as the (normalized) total reduction of the splitting criterion brought by that feature. We decide to use all flux density ratios as well as $F_{\text{z}}$ and $F_{\text{W}1}$.}
\end{table}

To determine which features to use for our analysis we run the random forest classification with all 13 available flux densities and flux density ratios. In Table.~\ref{table:feature_importance} we show how much information gain each feature gives relative to the others. We calculate these importances as the (normalized) total reduction of the splitting criterion. As visualized in Fig.~\ref{fig:all_color_color_plots} the different classes can be distinguished by their colors, we see this reflected in the feature importances: the flux density ratios lead to the most information gain for the random forest (see Table.~\ref{table:feature_importance}). The flux density ratios alone however remove the brightness information which would capture a luminosity evolution for different classes. To capture the brightness information we choose to use one flux density per survey. Since each flux density has approximately equal importance we choose the $\text{z}$ and $\text{W}1$ bands where our targeted quasars have the most reliable detections in Pan-STARRS and ALLWISE. Therefore, we choose to use all flux density ratios as well as $F_{\text{z}}$ and $F_{\text{W}1}$. 

\subsection{Class selection}\label{crossvaldiationsection}

\begin{table}[]
\centering
\begin{tabular}{lcc|cc}
\toprule
~ &\multicolumn{2}{c}{recall [\%]} & \multicolumn{2}{c}{precision [\%]}  \\\cline{2-3} \cline{4-5}
classes &  macro   &  high-z   &  macro & highz \\
\toprule 
all     & $74 \pm  5$ & $83 \pm 3$ & $78 \pm 4$ & $89 \pm 10$ \\
three   & $93 \pm 2$   & $79 \pm 5$  & $96 \pm 4$  & $88 \pm 12$  \\
binary  & $87 \pm 2$   & $74 \pm 5$  & $94 \pm 6$  & $88 \pm 12$ \\
\toprule
\end{tabular}
\caption{\label{table:classes_selection} To select the number of classes to use for the classification we compare the results when using all classes from Table~\ref{tab:classes} to using only three classes (high-z, other quasar, star) or binary classes (high-z or other). The macro values for precision and recall improve for fewer classes as expected but for the targeted high-z class there is no benefit so we will use the full set of classes. }
\end{table}

Our training data is labeled with the classes given in Table~\ref{tab:all_classes}. It is worth investigating if reformulating the problem as a binary problem (high-$z$  quasar vs other) or three class problem (high-$z$ quasar vs other quasar vs star) would improve our results. For each case, we run a 5 fold cross-validation with our random forest. We calculate a range of statistics summarized in Table \ref{table:classes_selection} with errors giving the standard deviation of the 5 runs. Precision and recall are calculated as defined in Section~\ref{sec:definitions}. The "macro" sub-column is the average over all classes weighted equally and the "high-z" sub-column is the precision and recall just for our targeted class. As we reduce the complexity of the classification problem by using fewer classes the macro metrics should get better since confusion between classes that we combine gets ignored. This is the case: for both the binary and three class problems the macro statistics are significantly better. However the recall and precision for the high-z class are not getting better, in fact, the recall in the binary case is actually lower with 1.6 sigma. We conclude that in our analysis we do not see an improvement from reducing the number of classes so we will use the full set of classes as defined in Table~\ref{tab:classes}.  

\subsection{Cross-validation}\label{crossvaldiationsection}

Under the assumption that the training sample represents the true distribution of objects on the sky and all sources are real, cross-validation of our training sample can predict the performance of the algorithm. This means that under the assumption that our training set contains a representative set of quasars, our definition of completeness is a reasonable measure for the fraction of all findable quasars that our random forest correctly identifies. 
However, we will overestimate the efficiency of our algorithm if we measure it with the precision because the number of M stars is even more dominating on the sky than in our training data and we are neglecting artifacts in the Pan-STARRS+WISE dataset. This is why we identify the precision as an upper limit for the efficiency and will use a different approach to get a more realistic estimate in Section~\ref{sec_eff}. 

Still, cross-validation lets us evaluate the strengths and weaknesses of the algorithm. We train the random forest classification and regression on a random subsample of 80\% of the training set and apply it to the remaining 20\%. We can then compare the predicted class and redshift with the true ones. By separating the data for training and testing, cross-validation avoids biased results from overfitting the data.

\begin{figure*}
\gridline{\fig{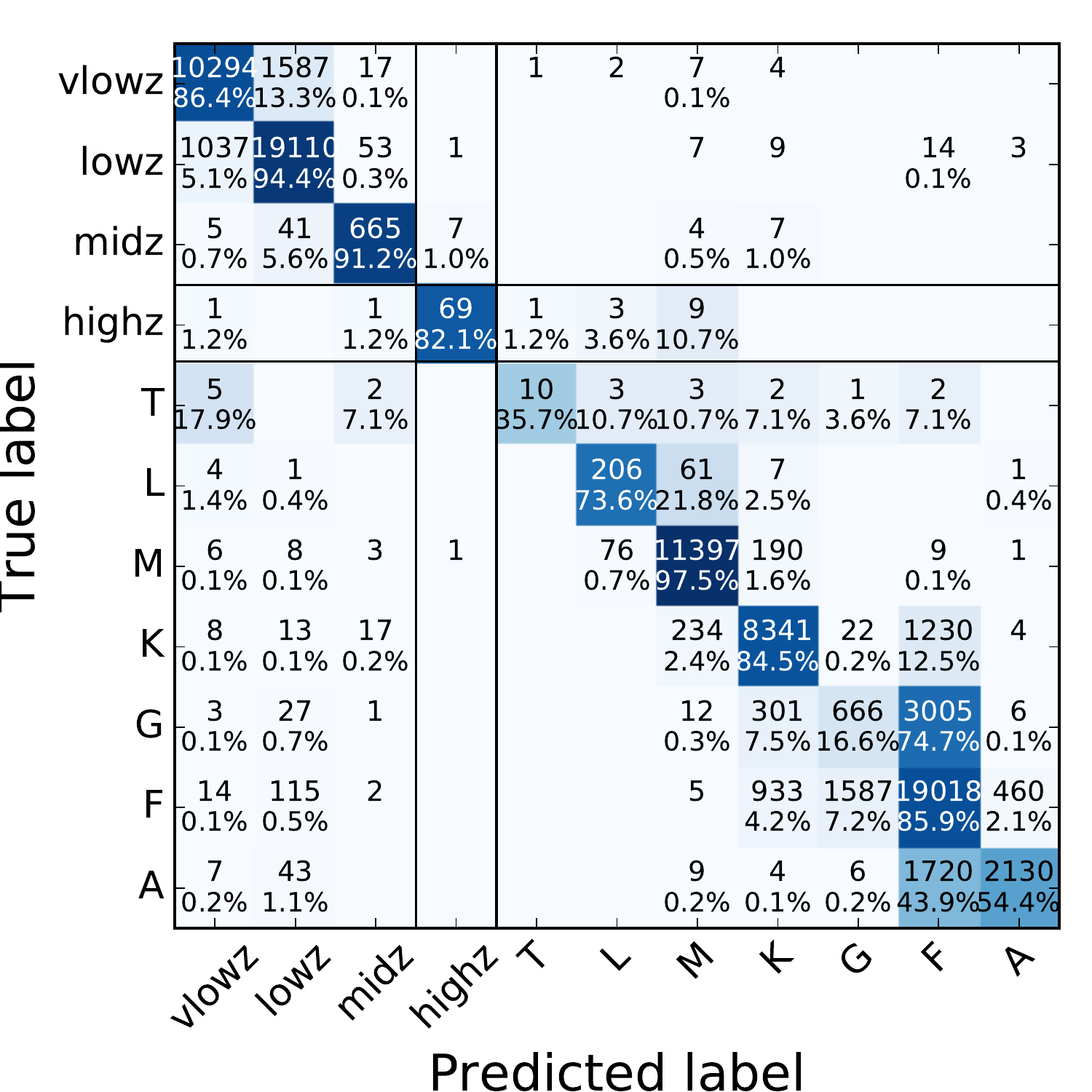}{0.63\textwidth}{(a)}
\fig{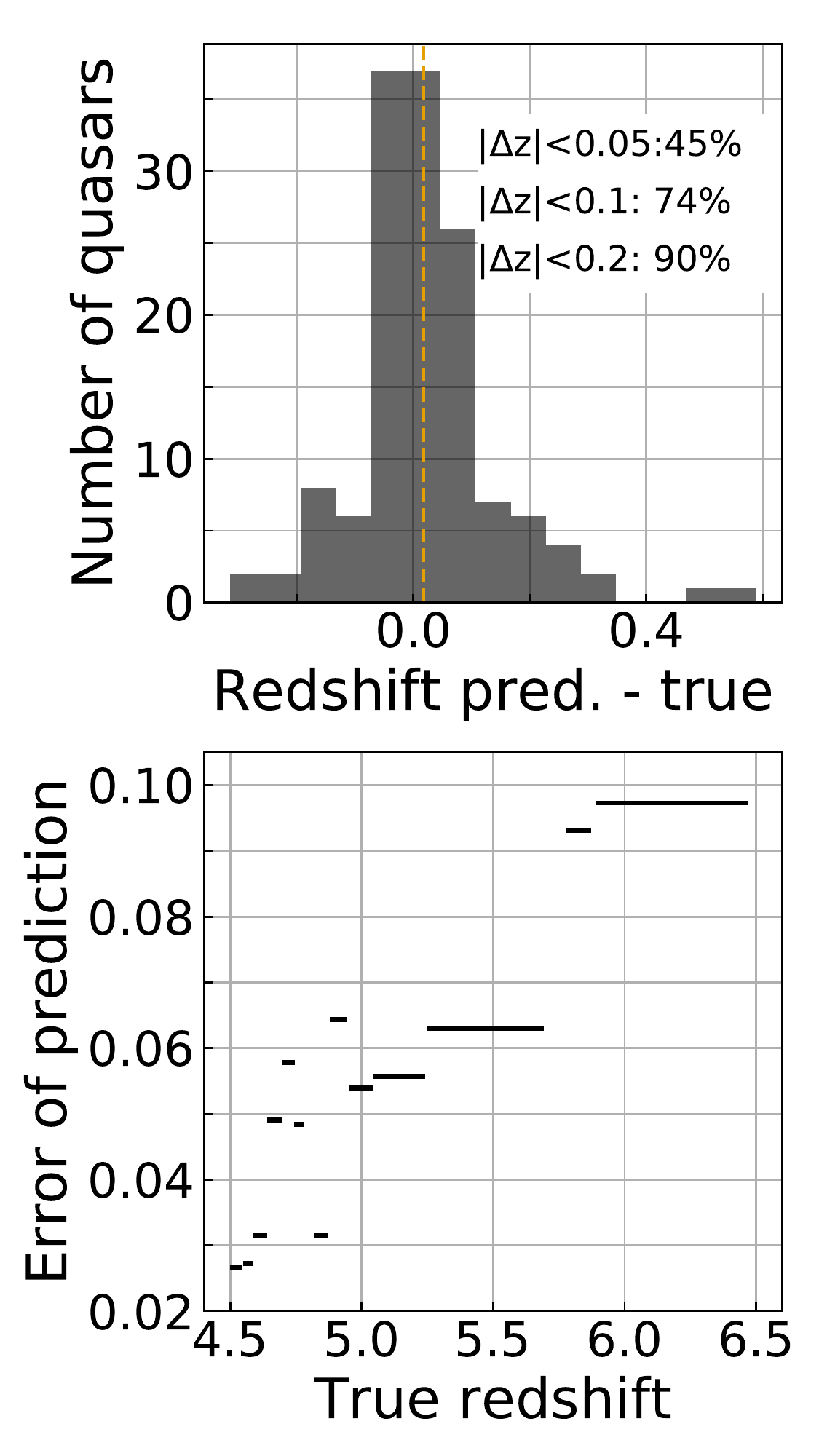}{0.35\textwidth}{(b)}
}
\caption{(a) Confusion matrix for the random forest classification on the cross-validation set. Overall the color information given to the random forest enables it to differentiate the classes for the majority of objects. For high-$z$ quasars, we identify the main contaminants as M, L and T dwarfs. The probabilities do not represent the performance of observations since the ratio of stars to quasars is underrepresented. \label{fig:confusion_matrix} (b) Cross-validation results for the photometric estimate of the redshift based on random forest regression for quasars with $z\geq 4.5$. The upper plot shows a histogram of the deviation between predicted and true redshift. The mean deviation, 0.015, is shown as an orange line. The lower plot shows the dependence on redshift as an absolute error vs redshift plot. The length of the line indicates the redshift bin used for averaging, chosen so that the number of objects per bin is constant.  \label{fig:regression_crossval}}
\end{figure*}


The random forest classification assigns each test object probabilities for each class. Later we can simply select our high-$z$ candidates by applying a threshold on the high-$z$ probability. First, however, we are interested in a comparison of all classes, so the most logical choice is to assign each object to the class with the highest probability. In Fig.~\ref{fig:confusion_matrix}~(a) we show the results of the classification on the cross-validation set in form of a confusion matrix. The matrix depicts how many objects of the true label class on the y-axis are classified to belong to the predicted label class on the x-axis. The majority of objects fall into the diagonal fields, demonstrating that our classifier assigned the correct label to them. The confusion between different types of stars is not concerning for our goal. Confusion between neighboring redshift bins is largely the result of objects right at the border between them and therefore also not concerning. As expected the most relevant contaminants for high-$z$ quasars are M, L, and T dwarfs, as the random forest classifies more than 15\% of high-$z$ quasars into those classes. In this case, there is only one star labeled as a high-$z$ quasar, but the balance between completeness and efficiency depends on how we define the cutoff probability for a high-$z$ classification. The highest probability approach chosen here, therefore, maximizes efficiency for lower completeness. Since M, L and T dwarfs far outnumber the high-$z$ quasars on the sky, the contamination will become significant at the redshift region of the strongest overlap in color space at around $z\approx 5.4$. Since the random forest regression will predict these objects around the same redshift, we will be able to exclude a large fraction of contaminants based on the regression by excluding highly contaminated redshift regions. A common approach to quantify the balance between completeness and efficiency is the ROC curve, which in our case is an almost perfect step function. The ROC score (area under the curve) for the high-z class vs the others is 0.99993.

%

We further analyze how accurate the random forest regression predicts the redshift for a cross-validation set of 20\%. We will differentiate two versions of the random forest regression. First, the \textit{full regression} where we train on quasars from the full range and can predict the redshift of any quasar. Second, the \textit{high redshift regression} where we only train with $z>4.5$ quasars and use it to predict the redshift of objects with class high-$z$. While the former covers a larger redshift range, the latter provides more accurate redshift estimates for the high-$z$ class candidates, because low redshift outliers in the \textit{full regression} training set can skew the result to lower redshifts.


The top of Fig.~\ref{fig:regression_crossval}~(b) shows the distribution of the difference between the predicted and true redshifts for the \textit{high redshift regression} when applied to the 20\% test set of the training data. 
90\% of objects have predictions within 0.2 of the true redshift. Since the algorithm can only find new quasars that look similar in color space to the training set, it is to be expected that this represents the performance of observations. As we will see in Section~\ref{section:observations_results} this accuracy is consistent with test observations. It should be noted that the accuracy of our redshift estimate is a strong function of the redshift. This is highlighted in Fig.~\ref{fig:regression_crossval}~(b) bottom. Here we show the absolute error of the prediction vs the true redshift. For this, we ran the prediction multiple times with different training set splits and then averaged the error over bins with equal amounts of objects. One outlier around redshift 4.7 was removed. We expected this increase of error with redshift because 1) there are fewer training objects at higher redshift and 2) higher redshift quasars appear fainter and thus have higher photometric uncertainties. We also note that especially towards the high redshift end the number of training and test objects becomes very small, so overfitting and redshift gaps in the training set can lead to significant additional inaccuracy in the redshift estimate when applying it to new data that is not captured in our cross-validation. The total training set for the \textit{high redshift regression} has 695 objects, only 50 are above redshift 6.
The \textit{full regression} applied to the same cross-validation set of $z\geq 4.5$ quasars gives similar results, but with a bias towards lower redshifts. 
In this case, only 69\% of cross-validation objects have predicted redshifts within 0.2 of the true redshift. In addition, the mean of the predicted redshifts is too low by $\delta z=-0.24$, because some objects are incorrectly predicted to be very low redshift quasars. Therefore, we decided to use the \textit{high redshift regression} for our candidate selection. 


\section{Estimating the selection efficiency}\label{sec_eff}

While the random forest approach returns a reasonable estimate for the completeness, the efficiency is overestimated due to the under-representation of contaminants in the training set. One way to deal with class imbalance is to use priors for the different classes as for example done in \citet{Bailer-Jones2019}, however the random forest approach we use here does not necessarily produce reliable probabilities even for the case of balanced classes which would be necessary \citep{2018arXiv181205792O}.

Therefore, we turn to a different approach to independently estimate the efficiency by exploiting the position information of our candidates which we have not used for the random forest. 
When averaging over large enough scales the distribution of stars on the sky is a function of galactic latitude, with more stars near the galactic plane. Quasars are more uniformly distributed over the sky, at least when averaging relatively large areas so small scale clustering averages out. So the idea is to estimate the distribution of target quasars and the dominant contaminants along the galactic latitude. This then allows us to estimate the efficiency by determining which combination of the two best recovers the distribution of our candidate set.

Any model of the stellar distribution on the sky will be dependent on the sensitivity limit of the survey and the stellar type. Therefore, we refrain from building a model of the stellar sky distribution but instead extract the distribution from our catalog data. The dominant contaminant for our targeted high-$z$ quasars are M stars (see Fig~\ref{fig:confusion_matrix} (a)). To have enough objects we make use of our random forest classification by taking 1 million objects that are predicted to be M stars (i.e. the M-star class has the highest probability). We note that this sample is not perfect and likely contains some artifacts, residual galaxies that were missed by our morphology cut and miss-classified quasars. Our cross-validation in Section~\ref{crossvaldiationsection} has however shown that the M star classification is quite reliable making this adequate for our purposes.
This allows us to estimate the distribution of contaminants of our selection (STARS). 
We model the quasar distribution (QUASARS) by uniformly sampling sources on a sphere, applying the same restrictions on the sky area.

We now calculate normalized histograms (h) as a function of galactic latitude for the candidate sample (CAND), the uniform distribution (QUASARS) and the M-stars (STARS) using the same bins in galactic latitude.
Now we assume that the distribution of quasar candidates can be modeled as a linear combination of the uniform distribution and the M-stars:
\begin{equation}
    h_{\text{CAND}, i} = \alpha \, h_{\text{QUASARS}, i} + (1-\alpha)\, h_{\text{STARS}, i}
\end{equation}
The suffix $i$ indicates the galactic latitude bins and $\alpha$ is the ratio of quasars to stars in our candidate set.
The efficiency of our candidate set is equivalent to the fraction of quasars to stars in our candidate set. Therefore, determining $\alpha$ provides a direct estimate of the efficiency of our candidate set.

%

\begin{figure}
\includegraphics[width=0.47\textwidth]{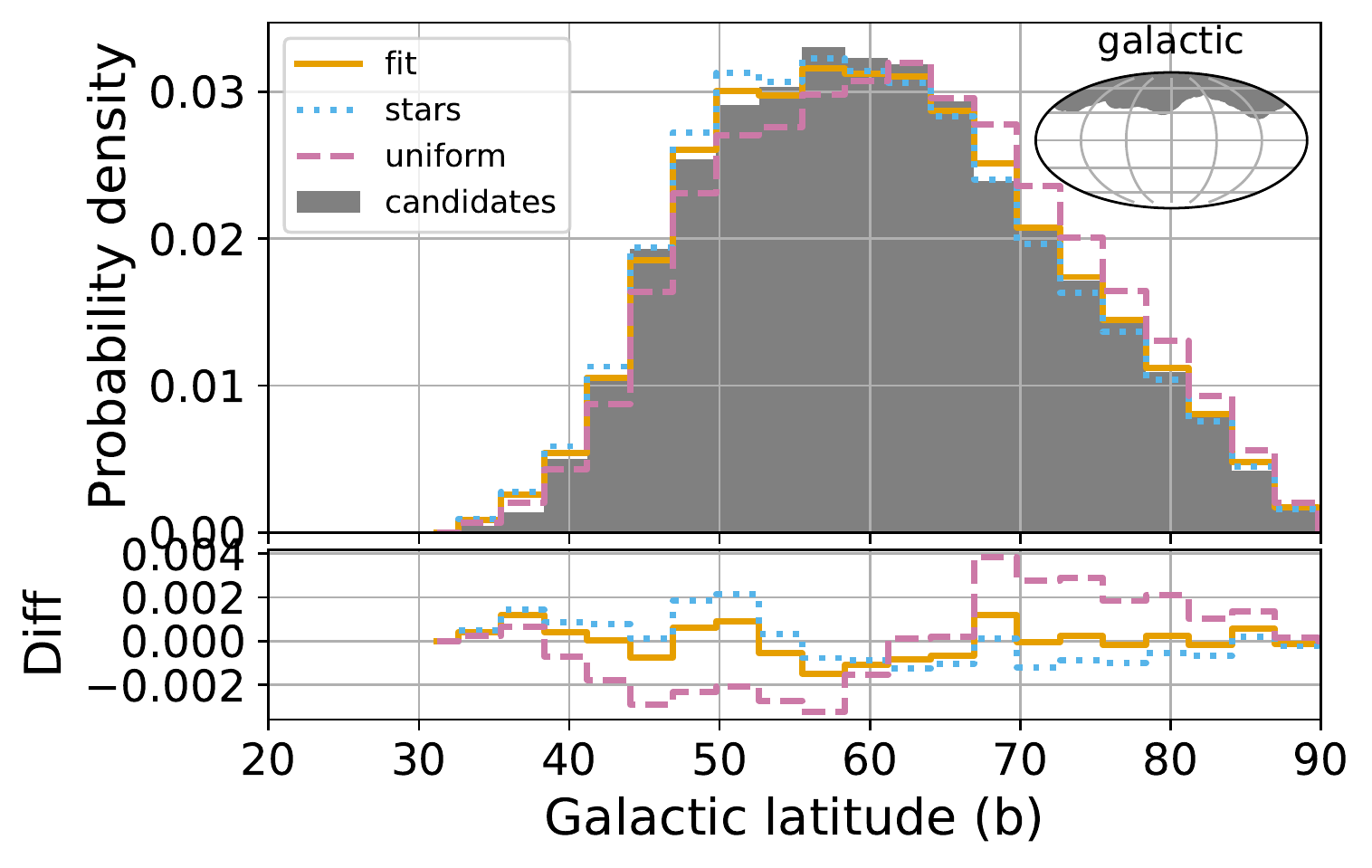}
\caption{\label{fig:efficiencytest_visualized}Test of the efficiency estimate for a subsample of the \citet{Richards2002} selection with 30.5 \% efficiency (number of spectroscopically confirmed quasars/number of candidates). Our estimate for this dataset gives $29.1^{+1.9}_{-3.4}\%$ for the efficiency, showing that our method works for this test case. The plot shows a normalized histogram of the candidates in Galactic latitude. The fit shown in orange is the weighted combination of the M-star (blue dots) and uniform (violet dashed) distributions, weighted by the efficiency. The lower plot shows the difference of the distributions to the candidate set. In the corner, a sky plot shows the area covered by our test sample in galactic coordinates. }
\end{figure}

To do this, we perform a minimization algorithm to find $\alpha$. In particular, we minimize the sum of the absolute differences between the left and right-hand side. Fig.~\ref{fig:efficiencytest_visualized} shows an example from our test of the method. For very large bin sizes there is no information content left since any slope gets averaged out. For very small bin sizes the quasars start to show measurable clustering, many bins of candidates are empty and the local depth variations in the survey do not average out anymore.

We sample a range of different bin sizes, randomly distributed between 20 and 100 bins and determine $\alpha$ for each realization. We quote the median of all determined quasar-to-star ratios and the 16th and 84th percentiles as the error. 
We implicitly assume that the estimates are independent of each other. For our test with SDSS data below we did not observe any concerning correlation. Still, this error only quantifies the statistical error. 

Our assumptions on the distribution of quasars and contaminants may introduce systematic errors.
We assume that the contaminants are mainly M stars, neglecting L and T dwarfs which might have slightly different distributions in our dataset. 
By construction, this method estimates the fraction of uniformly distributed objects, so it does not differentiate between quasars in our targeted redshift range and outside of it. Since in Section~\ref{crossvaldiationsection} we saw lower redshift quasars can be contaminants for our high-$z$ selection this has to be kept in mind. Furthermore, regions of high Galactic dust extinction may attenuate the quasar flux beyond our brightness requirements, making our quasar distribution dependent on dust and thereby dependent on galactic latitude. To minimize this effect we apply a cutoff in $E(B-V)$ for all selections as discussed below.

\begin{figure}
\includegraphics[width=0.47\textwidth]{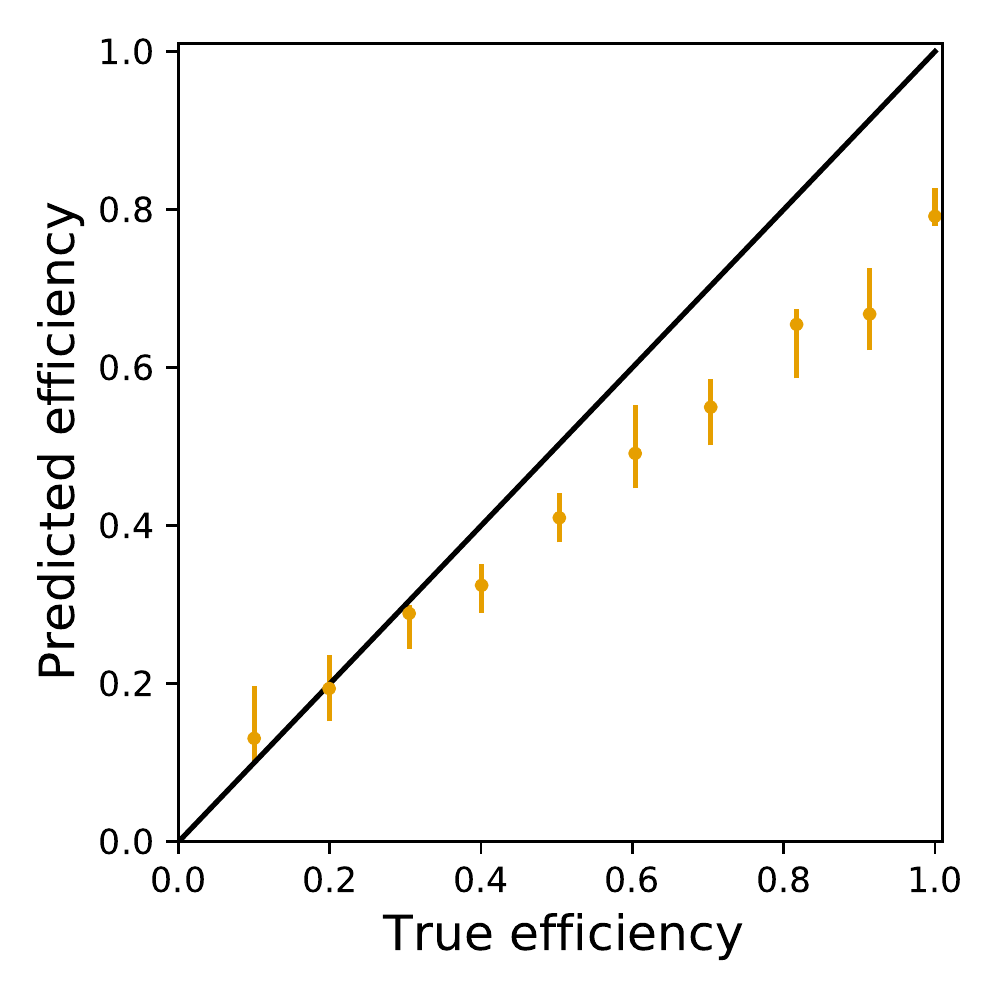}
\caption{Testing the efficiency estimate with SDSS spectroscopic data from \citet{Richards2002}. We combine samples of quasars and stars from the spectroscopic set to create datasets with a range of true efficiencies. The plot shows estimates for the efficiency from our method vs the true efficiency for these datasets. The black line indicates the correct result. The errors on the data-points are the 68\% confidence intervals which only capture the statistical error. This plot shows that there is also a systematic error but overall the method is working.\label{fig:eff_sdss}}
\end{figure}

We test our approach by showing that we can recover the efficiency of a set of quasar candidates where we have spectroscopic follow-up and therefore know the true efficiency. For this test, we use the original high redshift quasar selection from SDSS described in \citet{Richards2002}. This survey works well for our purposes since it was spectroscopically observed completely in a well-defined area. We simplify the footprint to $140< ra < 240$ and $0 < dec < 60$ where there is complete coverage. We also remove a suspicious region with a large over-density of objects (09h 00m 49s +47d 15m 34s with a radius of $5^\circ$) and apply a dust cutoff of $E(B-V)<0.1$. We take the objects classified as stars and the objects classified as quasars with $z>0.5$. This gives 15706 stars as well as 6889 quasars. Therefore, the true efficiency of this test dataset is $\frac{6889}{6889+15706}\approx30.5\%$. We note that this number is different from the published results since we applied a redshift cut for the quasars, ignored galaxies and only use a part of the observed area.
Now we take this test dataset and apply our efficiency estimation to it. As described above we compare the distribution of candidates vs galactic latitude with a uniform distribution and the distribution of our M-star sample. The best fit to the data gives an efficiency of $29.1^{+1.9}_{-3.4}\%$, with errors indicating the 68\% confidence interval. This shows that our method can recover the efficiency of the test dataset. The approach is visualized in Fig.~\ref{fig:efficiencytest_visualized} by showing the distributions for the candidates, M stars, a sample of uniformly sampled objects and our best fit. Since we are using a large number of sources the statistical error we give is relatively small. We note that using the M star distribution that we extracted from Pan-STARRS data to estimate the distribution of contaminants in the SDSS candidate set is a strong assumption and it is therefore quite surprising that the estimated efficiency matches the true value so well. This might indicate that our method can still give realistic results for the efficiency even when our modeling of the distribution of contaminants is quite rough.

To further test if our method also works for higher and lower efficiencies we take the SDSS test dataset from above and artificially create candidate sets with different ratios of quasars to stars. Specifically, we remove stars/quasars to increase/decrease the true efficiency of the test dataset, creating efficiencies between 10\% and 100\%. Then for each of these, we apply our estimate of the efficiency and compare it to the true value. Fig.~\ref{fig:eff_sdss} shows the results. Between true efficiencies of 10\% and 50\% our estimate is reasonably consistent with the correct value. For high efficiencies a systematic underestimation of the true efficiency is apparent. 
The accuracy at low efficiencies indicates that our star distribution is sufficiently similar to the stars in the selection. The deviation at high efficiencies indicates that the distribution of quasars in the selection is not quite consistent with our assumption of a uniform distribution. We identify two likely explanations for this behavior. It could be a physical difference, for example, small scale clustering of the quasars disturbing our result. The other possibility is that the selection was not made completely uniform. Spatial differences in the depth of the photometric survey data during the selection or in the follow-up observations may introduce these kinds of effects. 
For our analysis in this work, we are using a relatively conservative faint magnitude limit on the z band. This should ensure that the detection limit over the entire survey region is fainter than our requirement, giving relatively uniform coverage and thereby mitigating this issue. 

In summary, this method of estimating the efficiency of a quasar candidate selection is sensitive to any over-densities in the selection and non-uniformity, e.g. introduced by large scale variations in the survey depth. 
To avoid this when using the method on our high-$z$ candidate set below we check for and remove strong overdensities of candidates and make sure our targeted sky area is well defined.
Under these conditions, our test with the SDSS test dataset indicates that the method can predict the efficiency of a quasar candidate set up to a systematic error of less than 15\% between efficiencies of 20\% and 80\%.


\section{High-z candidate selection}\label{section:selection_results}
\subsection{Defining the selection}

We now apply the random forest classification and regression algorithms to our full Pan-STARRS+WISE photometric catalog data. 
To evaluate the completeness of our selection, we again split our training data into two parts one for training and one for evaluating the completeness. We decide to use the objects within two stripes $(ra \leq 60^\circ$ or $ra \geq 300)$ as well as ($-1.26 \leq dec \leq 1.26$) for the evaluation. This includes the Stripe 82 area which has been carefully surveyed for high redshift quasars and thus makes our completeness estimate more reliable \citep{McGreer2013}. Overall we use about $\sim22\%$ of the training set for evaluation and the rest to train the algorithm.


Our selection picks up larger numbers of candidates in regions of high Galactic extinction and near Andromeda. Therefore, we decide to apply additional restrictions on our photometric catalog data. We require a separation of at least 30$^\circ$ from the galactic center, a separation of 5$^\circ$ from Andromeda (0h 42m 44s +41d 16m 9s) and apply a dust extinction cut of $E_{B-V} < 0.1$.

After we removed the described areas our final photometric catalog includes around 59 million objects, covering 45\% of the sky\footnote{Sky coverage is measured as the fraction of a set of objects sampled uniformly on the sky that fulfill our restrictions on area and dust content.}. Table~\ref{tab:all_classes} shows the results of our random forest classification when assigning the class of the highest probability to the source.
\begin{table}[h]
\centering
\begin{tabular}{lr|lr}
A & 144,221 & T & 475\\
F & 12,786,096 & vlow-$z$ & 719,129 \\
G & 1,525,230 & low-$z$ & 973,060 \\
K & 17,709,898 & mid-$z$ & 1,217,562 \\
M & 26,648,920 & high-$z$ & 5,175 \\
L & 25,835 &  & 
\end{tabular}
\caption{\label{tab:all_classes}Random forest classification results for the Pan-STARRS+WISE catalog data of about 59 million point like objects that represents about 45\% of the sky for our brightness limits. The objects are assigned a class from Table~\ref{tab:classes} based on the highest probability. The number of quasars is likely inflated. Data artifacts and blended sources are not yet accounted for.}

\end{table}
As expected the predicted M-stars far outnumber our predicted quasars in the high-$z$ class. Yet, our training set over-represents high-$z$ quasars, therefore, we expect the number of good high-$z$ quasar candidates to be even lower. Similarly, since our training set under-represents L- and T-dwarfs in comparison to high-$z$ quasars, the number of predicted brown dwarfs is much less than high-$z$ quasars even though from observations we know it is the other way around.

\begin{figure*}
\includegraphics[width= 1. \textwidth]{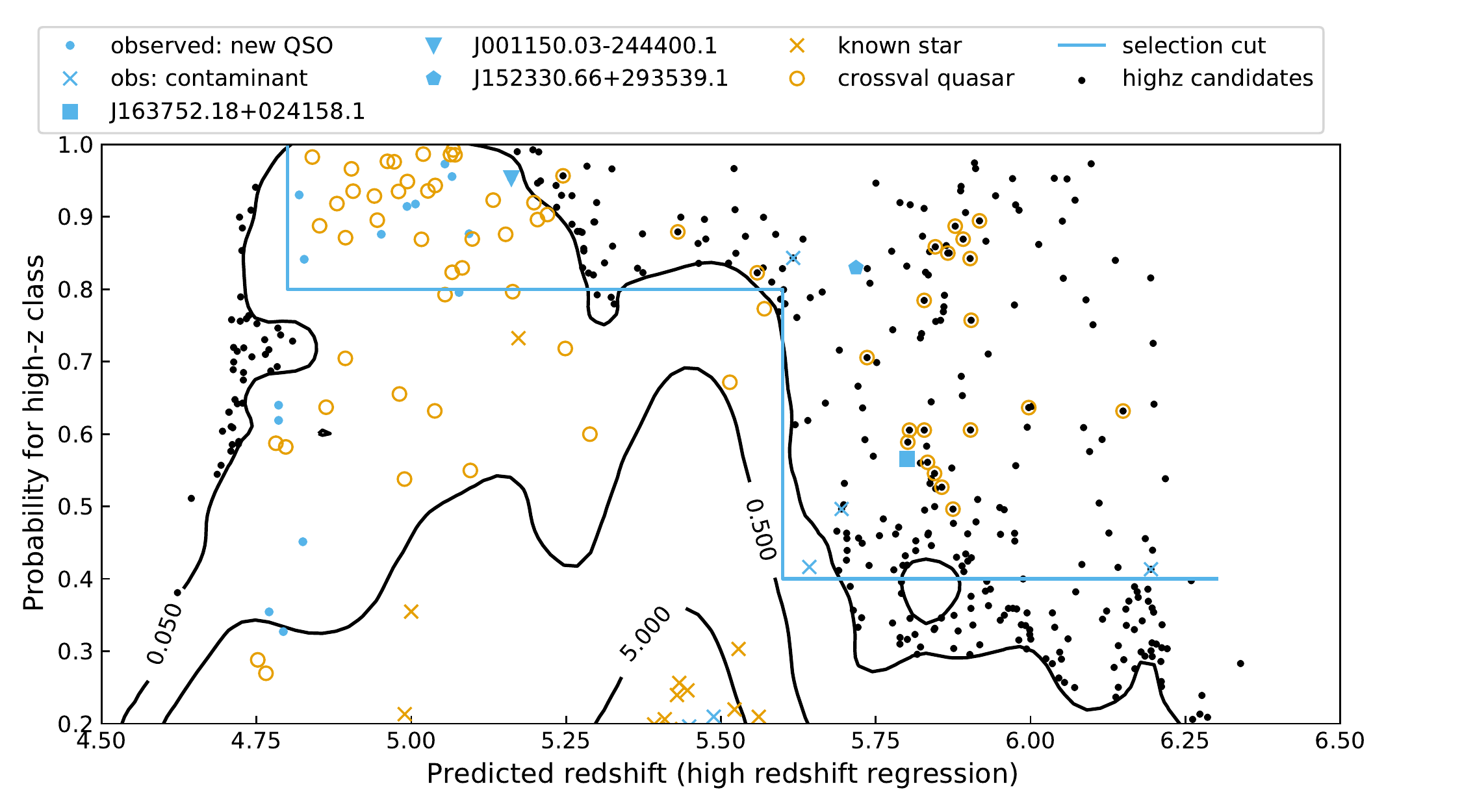}
\caption{Probability for the high-$z$ redshift class vs predicted redshift for our catalog data. The low limit is from the transition to the mid-z class and the high limit from our brightness requirement on the z band. The candidates, including known objects and artifacts, are shown in black. The majority is shown via a contour plot with 3 logarithmic density levels (the normalization of the numbers is arbitrary). The remaining objects are shown as black dots. Empty orange circles show the cross-validation quasars with which we estimate the completeness of the selection. Orange crosses show all known stars from the training set that are erroneously still in the selection. We show our \textit{high-$z$ candidate set} (Section~\ref{section:selection_results}) as a blue line. The observation results are shown as blue markers (Section~\ref{section:observations_results}), including preliminary ones not in the final selection. \label{fig:z_vs_high-z}}
\end{figure*}

\begin{figure}
\includegraphics[width=0.47\textwidth]{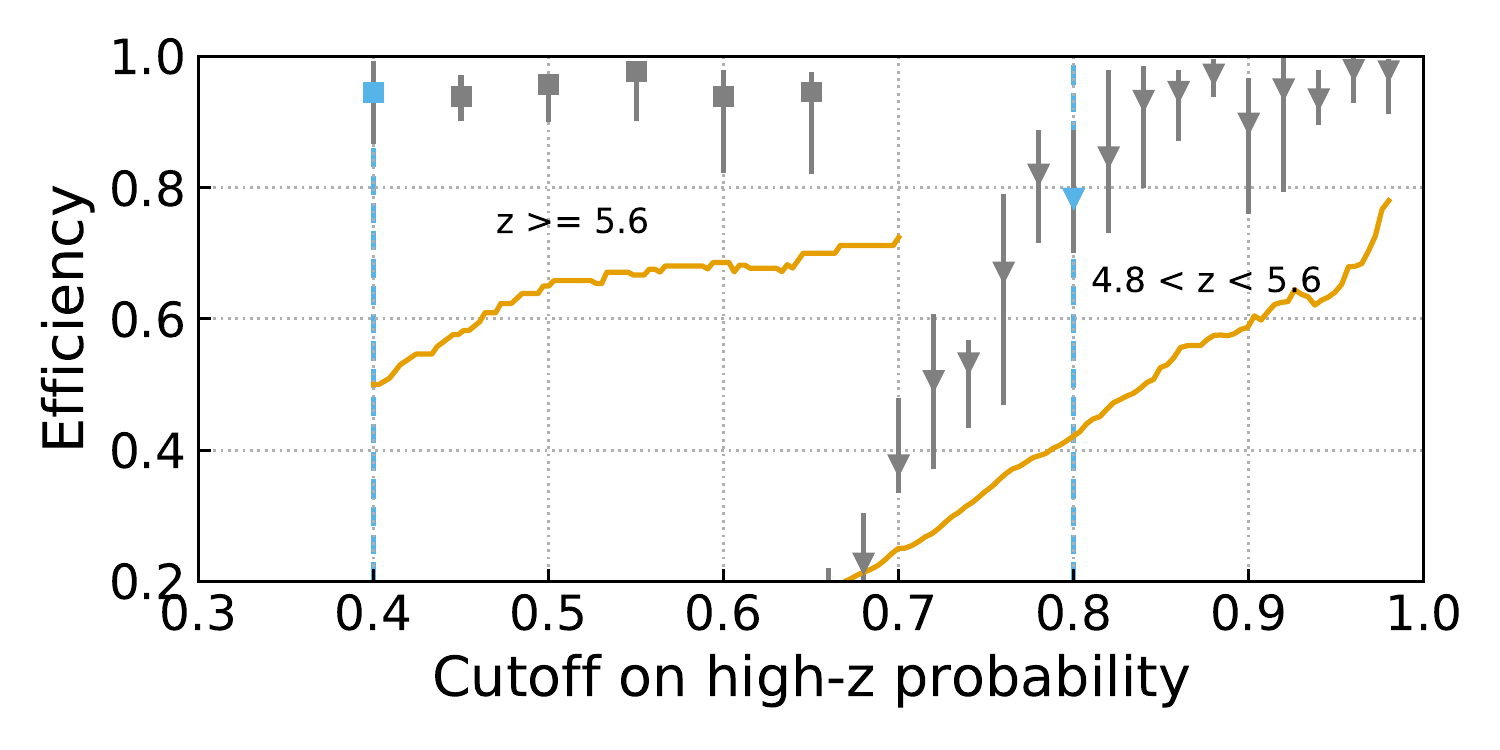}
\caption{The grey points show the estimated efficiency for different cutoffs on the high-$z$ probability. We separate the data into two ranges of predicted redshift: $4.8 < z \leq 5.6$ and $5.6 < z \leq 6.3$. Our choices are cutoffs of 0.8 and 0.4 respectively and are marked light blue, these cutoffs define the \textit{high-$z$ candidate set}. Since our candidate set contains known quasars we know that the efficiency is at least as big as the fraction of known quasars. We show this fraction of known quasars for different cutoffs as an orange line.  \label{fig:cand_selection}}
\end{figure}

The random forest classification algorithm provides us with a pseudo probability for each class. So far we simply assigned the class of highest probability, but now we instead look at the probability for high-$z$ quasar directly. Putting a cutoff on this pseudo probability lets us make a candidate selection where the cutoff can be tuned to our choice of efficiency vs completeness. While these pseudo probabilities provided by the random forest depend on the input training set and can not be trusted to represent an absolute measure, a higher high-$z$ class probability makes for a better quasar candidate. Therefore, we can improve the efficiency of our selection by increasing the cutoff on the high-$z$ class probability. At the same time, an increase in the cutoff will reduce the completeness since we are excluding more objects.

Fig.~\ref{fig:z_vs_high-z} shows the probability for the high-$z$ class vs the predicted redshift using the high redshift regression. We show the majority of candidates with a contour plot to visualize where the density of candidates is highest. To estimate the probability density we use a gaussian kernel density estimation applied to all candidates with high-$z$ probabilities above 15\%. We then show the probability density contours for 3 arbitrary density levels that are increasing by factors of 10. This way we can directly see that there is a large overdensity of candidates around redshift 5.4 and high-$z$ probabilities around 20\%. All candidates outside of the lowest probability contour are directly plotted as black dots.
At the low redshift edge, the number of objects with large high-$z$ probability drops off due to the transition from the high-$z$ to the mid-z class. We also find only few high-$z$ candidates beyond redshift $6.2$. This is expected as only one of our known $z\geq 6.3$ quasars passes our photometric requirements on the Pan-STARRS+WISE catalog data. In general, we expect a monotonous decrease in candidates with redshift since they are fainter.
We identify an overdensity of high-$z$ quasar candidates at $z\sim5.4$. There the trend of monotonous decrease in candidates is interrupted and towards low high-$z$ probability the number of candidates goes up much faster than at lower or higher redshift. There is no physical reason to expect much more quasars at that redshift, therefore we are likely seeing significant contamination from stars with similar colors. This is consistent with our evaluation of the cross-validation of the random forest classification (Section~\ref{crossvaldiationsection}): Around $z\sim5.4$ the contamination fraction rises, because of the photometric similarity between M stars and quasars at this redshift.

For the quasar candidate selection presented here, we have chosen to divide all candidates into two separate redshift ranges and treat them separately: $4.8 < z \leq 5.6$ and $5.6 < z \leq 6.3 $. 
Our choice is motivated by the sharp drop of candidate density around $z=5.6$. 
In our final selection, no cross-validation quasar is predicted to be in the wrong redshift range, allowing us to calculate the completeness for both sections separately.

We decide on the cutoff for the high-$z$ probability for each range by evaluating the efficiencies for a range of cutoffs. For our method of estimating the efficiency based on the sky distribution discussed in Section~\ref{sec_eff}, we need to first remove remaining artifacts and blended or extended sources. For this, we visually inspect image cutouts of the Pan-STARRS and WISE photometry for a manageable amount of objects. We inspect the objects with high-$z$ probability above 0.6/0.4 for the lower/higher redshift range. We remove an object if one of its Pan-STARRS images has an artifact interfering with the observation. We consider a nearby Pan-STARRS source detected in the z or y band as blended if it is within the $1\sigma$ radius of the PSF fit to the WISE source. We also remove all objects that are clearly extended in multiple bands of the Pan-STARRS imaging.

Then, we calculate the efficiency for a range of values to choose an optimum probability cutoff. The efficiency calculation follows Section~\ref{sec_eff} and uses the sky area of our Pan-STARRS+WISE catalog data. It covers about 45\% of the sky and is defined by:
\begin{itemize}
    \item $\rm{Decl.} > -30\,\deg$ 
    \item $|\rm{b}| \ge 20\,\deg$
    \item $> 30^\circ$ angular distance from the galactic center
    \item $> 5^\circ$ angular distance from Andromeda
    \item $E_{B-V} < 0.1$
\end{itemize}
The efficiency estimates for different cutoffs on the high-$z$ class probability are shown in Fig.~\ref{fig:cand_selection}. The uncertainties on the efficiency reflect the 50\% confidence interval. We expect lower values for the probability cutoff to include more contaminants resulting in a lower selection efficiency. 

This is exactly what we see in the lower redshift range ($4.8 < z \leq 5.6$) of our selection. 
The lower the probability cutoff is, the lower is our estimated efficiency. In this redshift range, the efficiency declines steeply for efficiency cutoffs below $\sim 80\% $. Therefore, we choose 80\% as our lower cutoff as indicated by the blue line in Fig.~\ref{fig:cand_selection}.

In the higher redshift range ($5.6 < z \leq 6.3$) much lower cutoffs still are predicted to have high efficiency. We choose the minimum lower limit tested: 40\%. We note that at first glance it seems like the efficiency drops for higher cutoffs, which is not expected. However, we argue that this just reflects the increase in uncertainty for higher cutoffs since there are only very few objects remaining. The full interval for each efficiency prediction in the higher redshift range is consistent with 1.   

The lower limit on the high-$z$ class probability is also indicated by the blue line in Fig.~\ref{fig:z_vs_high-z}. We retrieve a total of 617 candidates, of which we removed 102 during the visual inspection above (35 image artifacts, 42 blended sources, 25 extended sources). A total of seven known quasars are removed in that process. However, we do not relax our criteria on the visual inspection process, to only select candidates with highly reliable WISE photometry. Six of the seven known quasars we removed during visual inspection because they are blended in WISE. The seven known quasars removed during visual inspection represent 3.0\% of the known quasars in the candidate set, while overall 16\% of the candidate set is removed. We interpret this as an indication that the processing step is indeed reducing the fraction of contaminants in the final selection. 
 
In the end, we select a total of 515 promising quasar candidates which we call the \textit{high-$z$ candidate set}. 226 or $\sim43\%$ of these are already known quasars demonstrating the success of our selection method.

\subsection{Completeness and efficiency estimate}\label{sec:compl_eff_highz_selection}

For this sample of 515 quasar candidates, we now estimate the completeness and the efficiency. We estimate the completeness with the known quasars that we withheld from the training set. In particular, we define the completeness as the fraction of these known quasars that are in our final candidate set. The quoted uncertainties represent the 1-$\sigma$ confidence interval.

Since we are dealing with small sample sizes we use the Wilson interval to estimate the confidence interval for this binomial distribution, following the recommendation of \cite{Brown2001}. For large enough data-sets this converges back to the usual standard deviation of a Gaussian, for small data-sets it better captures the asymmetry in the error while retaining that the 1-$\sigma$ range captures 68.27\% of data points. 
We calculate a completeness of $66 \pm 7 \%$ for the redshift range of $4.8 < z \leq 5.6$ and a completeness of $83^{+6}_{-9}\%$ for the higher redshift range ($5.6 < z \leq 6.3$).


\begin{figure}
\includegraphics[width=0.47\textwidth]{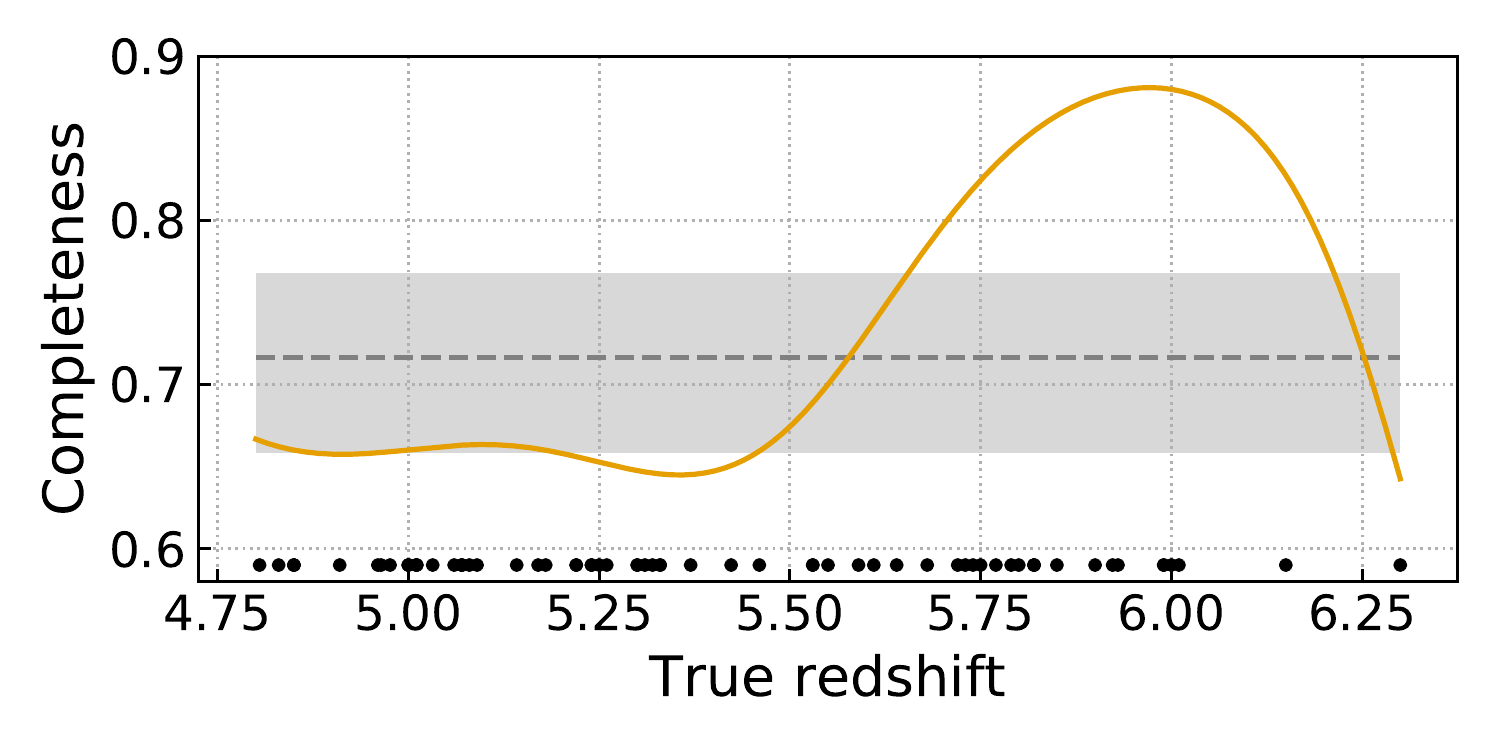}
\caption{
The solid orange line shows the completeness as a function of redshift for the \textit{high-$z$ candidate set} (Section\,\ref{sec:compl_eff_highz_selection}). 
We calculated two kde plots: both for the targeted known cross-validation quasars, one with only the ones still in the selection and the other for all of them. Dividing the two gives our estimate for the completeness. For each cross-validation quasar used, we show a black dot at its redshift at the bottom of the figure. The average completeness of all cross-validation quasars is $71^{+5}_{-6}\%$, shown as a dashed grey line with the 1 sigma error as a shaded box.
}
\label{fig:completeness}
\end{figure}
We show the completeness as a function of redshift in Fig.~\ref{fig:completeness}. To calculate this we applied a kernel-density estimate (kde) to both the targeted known cross-validation quasars remaining in the selection and in total. The ratio then gives our completeness estimate. For the kde, we used Gaussian kernels with equal weights for all points and bandwidths chosen with Scott's rule \citep{1992mde..book.....S}. Below redshifts of $z\approx5.6$ the completeness is nearly constant around a value of  $\sim67\%$. Above $z\approx5.6$ it rises to peak around $88\%$ at $z\approx5.9$. 
This behavior simply reflects that above predicted redshift $z=5.6$ we accept candidates with a lower high-$z$ class probability.
 
Above redshift 6 however, the completeness declines sharply. While this behavior is estimated based on only three cross-validation quasars with $z\ge6$, it signals that our method stops being effective at $z\ge 6$. Potentially, the small number of $z\ge 6$ quasars in our training set (52 total) might not allow for proper classification using the random forest method.

\begin{figure}
\includegraphics[width=0.47\textwidth]{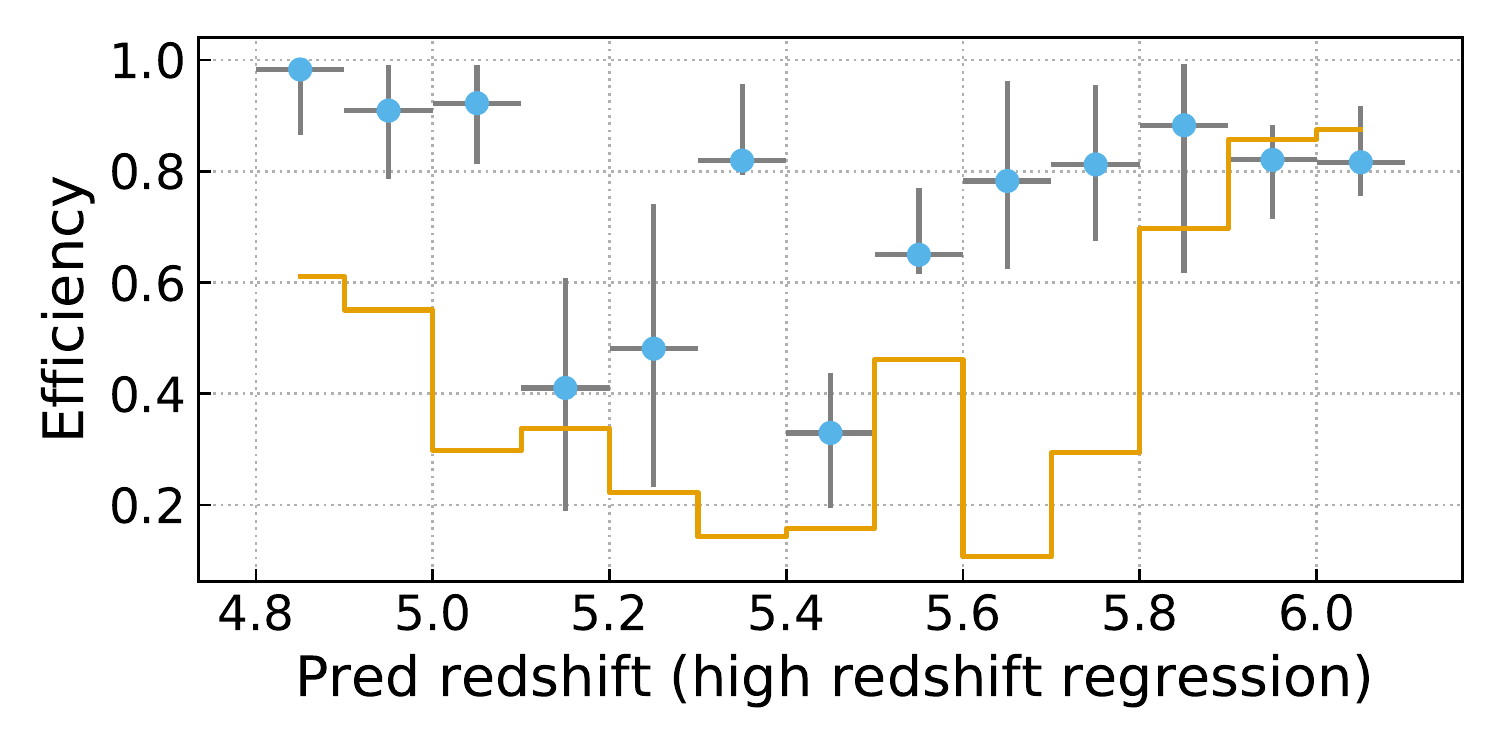}
\caption{The efficiency as a function of the predicted redshift for our \textit{high-$z$ candidate set} (Section\,\ref{section:selection_results}). The light blue data points show the median efficiency estimates based on our new methodology (Section\,\ref{sec_eff}). The redshift error bar depicts the redshift bin and the efficiency error is the 68\% confidence interval.
The orange line highlights the lower limit of the efficiency based on the known quasars in the selection. Where the efficiency estimate is above the lower limit we expect to find new quasars. \label{fig:efficiency}}
\end{figure}

We have fine-tuned our selection, in particular the lower limit on the high-$z$ class probability, to ensure high selection efficiencies. Indeed, our selection efficiency in the lower redshift range ($4.8 < z \leq 5.6$) is on average $78^{+10}_{-8}\%$ and for redshift of $5.6 < z \leq 6.3$ we reach $94^{+5}_{-8}\%$ (Fig.~\ref{fig:cand_selection}). The quoted uncertainties correspond to the 68\% confidence interval.
It is a good check for consistency that our method predicts efficiencies that are at least as high as the fraction of known quasars in our selection: The fraction of known quasars in the final set is 42\% and 50\% in the lower and higher redshift range, respectively.

Next, we estimate the redshift dependence of the efficiency of our candidate set. The efficiency is the fraction of our candidates that are actually quasars. To estimate this we again use our method from Section~\ref{sec_eff}.
We calculate the efficiency for bins with a width of 0.1 in redshift. Fig~\ref{fig:efficiency} shows our selection efficiency as a function of redshift. Part of our candidate set are known quasars, which tells us that the efficiency of the selection is at least as large as the fraction of known quasars in that bin. We show this minimum efficiency in orange. Whenever the lower efficiency limit and our estimated efficiency agree we do not expect to find new quasars in that redshift bin. When the estimated efficiency is larger we do expect the candidate set to contain quasars that are not yet known. 
In the redshift bin of $z=5.4-5.5$ we see the selection efficiency drop to the lowest value in our entire redshift range. This is likely the result of the significant overlap in color space of quasars with M stars at $z\approx5.4$ as we discussed in Section~\ref{crossvaldiationsection}.
Based on our efficiency estimate we expect to find the highest redshift quasars with this selection around $5.5 < z < 5.8$, where our predicted selection efficiency is above the lower limit.
At $z\ge5.8$ the efficiency prediction and the lower limit are consistent with each other. Therefore, we expect to find few new quasars at $z\ge5.8$. 
Finally, at the low end of our targeted redshift range, we also expect new quasars since the efficiency estimate is well above the lower limit.

Based on the estimate of the efficiency we can predict that our high-$z$ candidate sample contains $319^{+41}_{-33}$ quasars at $4.8 < z \leq 5.6$ and $100^{+5}_{-8}$ at $5.6 < z \leq 6.3 $. 
Subtracting the known quasars, we expect that our candidate set contains $148^{+41}_{-33}$ and $45^{+5}_{-8}$ new quasars in the lower and higher redshift range respectively, where the error is a 68\% confidence interval. 
Table~\ref{table:highzselection} summarises our predictions for the selection. We deliver the paper with a data-file containing the full \textit{high-$z$ candidate set}. Table~\ref{table:highzselection_description} describes the columns of the data-file.

\begin{table}
\ra{1.3}
\centering
\begin{tabular}{lcc} 
\toprule
Redshift range       & $4.8 < z \leq 5.6$  & $5.6 < z \leq 6.3 $ \\
Number of candidates & 409                 & 106                 \\
Completeness         & $66 \pm 7 \%$       & $83^{+6}_{-9}\%$          \\
Efficiency           & $78^{+10}_{-8}\%$   & $94^{+5}_{-8}\%$    \\
Known quasars        & 171                 & 55                  \\
Predicted new quasars          & $148^{+41}_{-33}$ & $45^{+5}_{-8}$ \\
\toprule
\end{tabular}
\caption{\label{table:highzselection} Summary for the \textit{high-$z$ candidate set}. The calculation of these properties is discussed in Section~\ref{section:selection_results}. All errors give 68\% confidence intervals.}
\end{table}


\begin{table}[!htb]
\ra{1.3}
\centering
\begin{tabular}{lp{42mm}} 
\toprule
Column name       & Description \\
\toprule
WISEDesignation & Name in wise catalog \\
RAdeg & Ra in Pan-STARRS catalog \\
DEdeg & Dec in Pan-STARRS catalog \\
zPSFStackMag & z stacked PSF magnitude \newline in Pan-STARRS \\
HighzProb & Probability for high-$z$ class \\
QsoProb & Summed probability for \newline quasar classes \\
MstarProb & probability for M star class \\
PredictedRedshift & \textit{high redshift regression} result \\
SpectroscopicRedshift & Redshift determined from \newline spectrum \\
KnownQuasar & Boolean whether quasar is \newline known in literature \\
PhotometricFollowUp & Boolean whether we obtained photometric follow-up \\
Observed & Boolean whether we took a spectrum of the object \\
StillToObserve & Boolean whether object still has to be observed \\
\toprule
\end{tabular}
\tablecomments{Only a portion of this table is shown here to demonstrate its form and content. It is published in its entirety in the machine-readable format.}
\caption{\label{table:highzselection_description} List of columns of the  \textit{high-$z$ candidate set}.}
\end{table}

\section{Observations}\label{section:observations_results}

During the development of our selection process, we have followed up some of our quasar candidates with photometry (6) and spectroscopy (37). 

Photometric follow-up observations have been performed with the Nordic Optical Telescope (NOT) using the NOT near-infrared Camera and spectrograph \citep[NOTCam; ][]{Abbott2000}. The observations were taken on 2019 May 17 to 20. We used the OB generator for scripting. For our observations in the J band, we used 9 point dithering. We read out the detector in ramp-sampling mode with 9 seconds between readouts, a total of 10 times. Giving us an effective exposure time of 90 seconds for each of the 9 pointings. Depending on seeing and brightness of the object we executed this 1, 2 or 3 times to get enough signal to noise to measure the magnitude. 

Additionally, we were able to secure optical spectroscopy with the Goodman High Throughput Spectrograph \citep[HTS; ][]{Clemens2004} on the Southern Astrophysical Research Telescope (SOAR), with MODS on the Large Binocular Telescope (LBT)  \citep{Pogge2010}, with the Magellan Baade telescope's Folded port InfraRed Echellette \citep[FIRE;][]{Simcoe2013}, and with FORS2 on the Very Large Telescope (VLT).

Spectra with Goodman HTS on SOAR were taken using the 400\,g/mm grating with a central wavelength of $7300\,\text{\AA}$ resulting in spectra with a wavelength coverage of  $\sim5300-9300\,\text{\AA}$ (GG-495 blocking filter). All observations used the red camera in 2x2 spectral binning mode. For the $\sim5300-9300\,\text{\AA}$ "red" spectrum we exposed for $900\,\rm{s}$ using the $1\farcs0$ slit, which provides a resolution of $R\approx830$. The spectra were reduced with IRAF.
We took FIRE high-throughput prism spectra using the $1\farcs00$ slit over the spectral range of $\sim8250-25200\AA$ with a resolution of $R=300-500$.
For the LBT we used MODS in the red channel only mode. We use the G670L grating with blocking filter GG495, a slitwidth of $1\farcs20$ and an exposure time of 1200sec.
Spectra with FORS2 on the VLT were taken using the GRIS\_600z+23 grism with the OG590+32 filter, a slitwidth of $1\farcs30$ and an exposure time of 900sec.

We present 20 newly discovered high redshift quasars and discuss them in the context of the \textit{high-$z$ candidate set} presented in Section\,\ref{section:selection_results}. However, some candidates were selected before we finalized our candidate selection methodology. We provide information on their original selection where appropriate.\\

\subsection{$z>5.6$ High-z Candidate Follow-up}

We select a subset of our final high-$z$ candidate catalog and require all candidates to have prediction redshifts of $z_{\rm{RF}}>5.6$ in both the \textit{full regression}, using training quasars at all redshifts, and the \textit{high redshift regression}, using training quasars at $z>4.5$.
We retain 59 promising quasar candidates, which nominally have a selection efficiency of $86^{+11}_{-34}\%$ as estimated by our new method (see Section~\ref{sec_eff}). 
32 of the candidates are already known quasars at the time of our selection, leaving 27 unknown objects. 
From our efficiency estimate, we expect through error propagation that for these 27 objects our success rate to find quasars should be $69^{+24}_{-52}\%$ with the 68\% confidence interval as the error. We can compare this to a naive estimate based on Equation~\ref{def_precision} which would lead us to predict 100\% for the efficiency since there is no known star from our test set in our final selection (see Figure~\ref{fig:z_vs_high-z}). 


One of these candidates, J112143.62-071839.4, was recently identified by \citet{Yang2019} as a z=5.71 quasar. 

\begin{figure}
\plotone{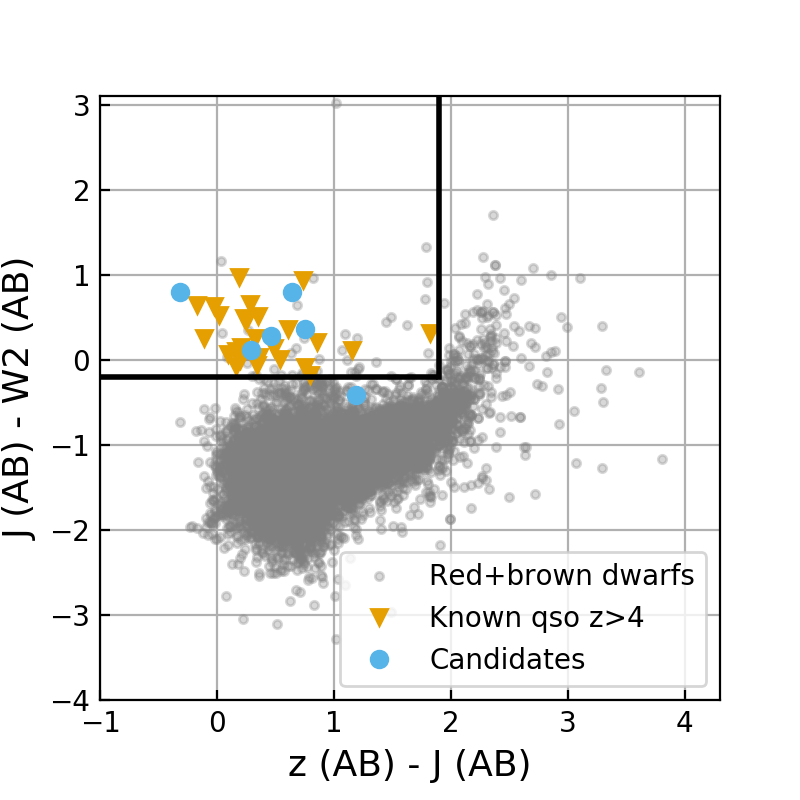}
\caption{Color-color diagram for observations of 6 quasar candidates. The blue dots are the candidates where we observed the J band with NOT (see Table~\ref{tab:lapalma}), 5 out of 6 are consistent with known quasars at this redshift (orange triangles). In grey, a sample of contaminant red and brown dwarfs is shown. For the known objects we only plot the brightest objects where 2MASS gives a measurement of the J-band. The black line gives a reasonable cut for a quasar selection.
\label{fig:lapalma}}
\end{figure}
\begin{table}
\ra{1.3}
\centering
\begin{tabular}{lll}
\toprule
wise\_designation   & J band (VEGA)   & promising \\
\toprule
J124359.84+173445.3 & $18.59\pm 0.06$ & Yes                   \\
J140531.13+735243.8 & $18.45\pm 0.03$ & Yes                   \\
J145836.16+101249.5 & $18.27\pm 0.02$ & Yes                   \\
J145950.96-181251.7 & $18.07\pm 0.05$ & No                    \\
J152055.71+431652.4 & $19.09\pm 0.04$ & Yes                   \\
J152330.66+293539.1 & $19.61\pm 0.13$ & Yes                  \\
\toprule
\end{tabular}
\caption{Results of the photometric follow-up observations. The J band measurements are calibrated with the 2MASS sources found in the field of view of our observations. 5 out of 6 objects have J band magnitudes consistent with high redshift quasars, i.e. they fulfill our color cut in Fig.~\ref{fig:lapalma} and are therefore promising.\label{tab:lapalma}}
\end{table}
The J-band is very effective in differentiating the classes since red and brown dwarfs tend to have more flux in their spectrum for this band than otherwise similar-looking quasars. 
We followed-up 6 candidates with J-band photometry using NOTCam. The results are summarized in Table~\ref{tab:lapalma}.

%
Fig.~\ref{fig:lapalma} shows the zJ-JW2 color-color diagram combing the Pan-STARRS magnitudes with the J band from our NOTCam follow-up observations (blue dots). As a point of comparison, we also plot our known objects for which J band information from 2MASS is available \citep{2006AJ....131.1163S}\footnote{We obtained the ALLWISE data from IRSA (see Section~\ref{subsection:training_set}) where it is already cross-matched with 2MASS.}. We note that 2MASS is shallower than our follow-up observations. The known quasars and stars have mean J band AB magnitudes of 17.6 and 16.5 while our 6 follow-up observations have a mean of 19.6. However, to first-order quasars at the same redshift have similar colors with only minor luminosity evolution. 
Promising quasar candidates can be separated from likely dwarf stars with a color cut shown in black ($\text{z}-\text{J} < 1.9$ and $\text{J}-\text{W}2 > -0.1$, both in AB mag). Five out of our six observed objects make the color cut. A close-by source is evident in the J-band photometry of our only non-promising candidate, J145950.96-181251.7. Therefore, it is likely blended in WISE, which could explain the false classification. Overall our NOT photometric follow-up observations indicate that our candidate set does contain promising candidates. 

Furthermore, we were able to obtain five follow-up spectra of our selection. These objects were not prioritized by the high-$z$ probability, we observed the objects in the candidate set with the best visibility at the observatories. We identified three objects as contaminants and two as quasars at $z\sim5.7$. 

Counting the \citet{Yang2019} quasar and the likely quasar at lower redshift the selection efficiency would be 3 out of 6 or 50\%. Since we did not observe the sixth object, a more conservative counting would be 2 new quasars in the targeted redshift range out of 5 observed or a selection efficiency of 40\%. From a naive approach we would have expected an efficiency of 100\% while our method for estimating efficiency predicted $69^{+24}_{-52}\%$. While our very small sample size does not allow conclusions about the accuracy of our approach we do argue that our method does give more realistic results that are consistent with our small test observation. 

In the following, we discuss the two newly discovered quasars individually. 
We note that we observed a sixth object: J032615.68-061358.2. The continuum looks like a power law, however, we do not see a Lyman-$\alpha$ break in our spectral range. This indicates that it likely is a quasar but at $z<5.4$ where the Lyman-$\alpha$ line moves out of our spectral range. The predicted redshift of $z=5.62$ was too high. We do not consider it for our efficiency test here because we can not confirm the classification. In Fig.~\ref{fig:z_vs_high-z} we show the object with a cross. 



\begin{table*}[!htb]
\ra{1.3}
\centering
\begin{tabular}{llllllll}
\toprule
WISE designation    & PS Mean ra  & PS Mean dec  & z Mag & M1450 & Tel/Instr & Obs. date & z                       \\
                    & (deg)       & (deg)        & (AB)  & (AB) &                      & (YYMMDD)  &                         \\
\toprule
J000425.84-211054.2 & 1.10781106  & -21.18168195 & 19.57820624               & -26.77032395              & SOAR/G HTS      & 180604    & 5.09                                    \\
J001150.03-244400.1 & 2.9585218   & -24.7333892  & 19.30589762               & -27.45251895              & SOAR/G HTS      & 180604    & 5.41                                     \\
J012947.32-295235.1 & 22.44703224 & -29.87629848 & 19.51007577               & -26.80463644              & SOAR/G HTS      & 180603    & 4.83                                     \\
J013539.29-212628.4 & 23.9137294  & -21.44122046 & 17.84342435               & -28.21850479              & SOAR/G HTS      & 180603    & 4.91                                     \\
J084347.77-253155.8 & 130.9490235 & -25.53213628 & 18.49298484               & -27.34917379              & SOAR/G HTS      & 180404    & 4.72                                    \\
J085943.27-003613.2 & 134.9301648 & -0.60363371  & 20.32148047               & -25.7887862               & SOAR/G HTS      & 180406    & 5.03                                    \\
J093032.56-221207.5 & 142.6357036 & -22.20214902 & 18.09687224               & -27.98936448              & SOAR/G HTS      & 180406    & 4.86                                     \\
J094135.48-061547.0 & 145.39785   & -6.26308714  & 19.29794439               & -26.97598817              & SOAR/G HTS      & 180404    & 5.05                                     \\
J094418.13-200106.4 & 146.0756187 & -20.01850398 & 19.05723975               & -27.20861384              & SOAR/G HTS      & 180604    & 4.93                                     \\
J095139.70-274210.7 & 147.9153819 & -27.70348097 & 18.39840532               & -27.63989338              & SOAR/G HTS      & 180406    & 4.8                                      \\
J100451.83-091751.7 & 151.2159282 & -9.29779768  & 19.22792522               & -26.8103096               & SOAR/G HTS      & 180604    & 4.91                                     \\
J103020.14-042105.7 & 157.583914  & -4.3515849   & 18.88467506               & -27.02610394              & SOAR/G HTS      & 180404    & 4.66                                     \\
J105541.85-103007.6 & 163.9243208 & -10.50207368 & 19.99007524               & -26.51733591              & SOAR/G HTS      & 180406    & 5.04                                     \\
J110942.97-285521.0 & 167.428771944 & -28.9223126129 & 20.035114               & -26.02474743              & VLT/FORS2      & 210404    & 5.01                                     \\
J141359.37-212713.7 & 213.4974405 & -21.45382469 & 20.30688036               & -25.75043534              & SOAR/G HTS      & 180406    & 4.92                                     \\
J142829.63-213059.9 & 217.1233865 & -21.51677998 & 20.04845089               & -25.97715567              & SOAR/G HTS      & 180406    & 4.87                                     \\
J150542.94-071718.1 & 226.4290075 & -7.28845091  & 20.16459203               & -26.11565751              & SOAR/G HTS      & 180406    & 4.99                                     \\
J152330.66+293539.1 & 230.8777384 & 29.5943535   & 20.17168355               & -26.446892                & LBT/MODS             & 190611    & 5.73                                     \\
J163752.18+024158.1 & 249.4674059 & 2.6995546    & 19.22674035               & -27.09365806              & SOAR/G HTS      & 180602    & 5.76                                     \\
J232952.78-200039.1 & 352.4699164 & -20.01088649 & 18.43736995               & -27.82847382              & SOAR/G HTS      & 180603    & 5.03                                     \\ 
\toprule
\end{tabular}
\tablecomments{Table 7 is also available in the machine-readable format with spaces in column names removed.}
\caption{\label{table:observedobjects} List of the newly discovered quasars reported in this work. The listed z band magnitude is based on the PSF stacked magnitude from Pan-STARRS and corrected for extinction. The listed redshift is estimated from the Lyman-$\alpha$ emission. These spectroscopic redshifts are accurate to about $\Delta z=0.05$. G HTS is short for GoodmanHTS.}
\end{table*}

\subsection*{J152330.66+293539.1 - $z = 5.73$}\label{sec:quasarmissedbycolorcut}
J152330.66+293539.1 is a newly discovered quasar at redshift 5.73 based on Lyman-$\alpha$ emission. The predicted redshift was 5.72 and the high-$z$ probability was 0.83. The object was part of our NOT photometric follow-up observations, where we measured the J-band magnitude. The obtained colors where $\text{J}-\text{W}2 = 0.80$ and $\text{z}-\text{J} = -0.32$ in AB which are consistent with quasars at this redshift as seen in Fig.~\ref{fig:lapalma}. We observed this object with the MODS spectrograph on the LBT and present the spectrum in Fig.~\ref{fig:quasar_spectra} together with the other discovered quasars. 

This object has a relatively blue color of $\text{i}-\text{z}=1.84$, so it would not be part of a typical color cut selection like \citet{Banados2016} where candidates were cut at $\text{i}-\text{z}>2$. This indicates that our method can find quasars that are missed by traditional color cuts even if our random forest is trained with objects largely from these selections. Our full \textit{high-$z$ candidate set} that we publish with this work contains a further 37 candidates with predicted redshift above 5.6 that do not fulfill this color cut. There are only 7 known quasars above redshift 5.6 that do not fulfill the color cut.

\subsection*{J163752.18+024158.1 - $z = 5.76$}

J163752.18+024158.1 is a newly discovered redshift 5.76 quasar based on the Lyman-$\alpha$ emission. 

The redshift prediction in our high-$z$ candidate catalog $z=5.80$ (\textit{high redshift regression}) is very close to the observed redshift and the high-$z$ probability was 0.57.
We observed this object with the GoodmanHTS on the SOAR telescope.

\subsection{$z=4.6-5.4$ High-z Candidate Follow-up}

We have tested our random forest selection method with pilot observations during the development process.
We used a preliminary version of the algorithm to make a selection targeted at redshift 4.8 to 5.4 and obtained 31 optical spectra, out of which we successfully identified 17 new quasars. 
Eight of these quasars are retained within our final \textit{high-$z$ candidate set}, but none of the contaminant objects are selected anymore. This indicates that our selection improved in robustness. The newly discovered quasars, which did not make it into our final selection (a total of nine) are either at lower redshift than our targeted selection (three have observed $z<=4.8$) or are very close to the redshift boundary. A total of eight have observed redshifts below $z=4.92$ and hence their classification shifted towards the mid-z class.
Another newly discovered quasar at $z=5.03$ just barely missed the 80\% cutoff on the high-$z$ probability (79.6\%) and thus was excluded from our final candidate list. 

These preliminary observations were already very successful with 55\% of observed candidates being newly discovered quasars. With the improvements to our selection discussed above the final selection is expected to be even better in this redshift range around $z\approx5$. 

Furthermore we were able to obtain follow-up observations for one additional object in the lower redshift range of our final \textit{high-z candidate set} using FORS2. We identify the object, J110942.97-285521.0, as a quasar at z=5.01. The predicted redshift was 5.09. In the following, we discuss the discovered quasars.

\subsubsection*{J001150.03-244400.1 -- $z = 5.41$}

We discovered J001150.03-244400.1 at redshift $z=5.41$. While we selected and observed this object based on the preliminary selection described above, this quasar is also part of our final \textit{high-$z$ candidate set}. This quasar was observed with the GoodmanHTS spectrograph. We show the discovery spectrum in Fig.~\ref{fig:quasar_spectra} and list the object information in Table\,\ref{table:observedobjects}.
The random forest regression predicted a redshift of $z=5.16$, significantly lower than its real redshift. The high-$z$ probability was 0.95. Interestingly, the spectrum shows a strong absorption through in the Lyman-$\alpha$ forest at observed wavelengths of $6300-6500\textrm{\AA}$.

\subsubsection*{Seventeen new quasars at $4.6 \leq z \leq 5.1$}

\begin{figure*}
    \centering
    \includegraphics[width=\textwidth]{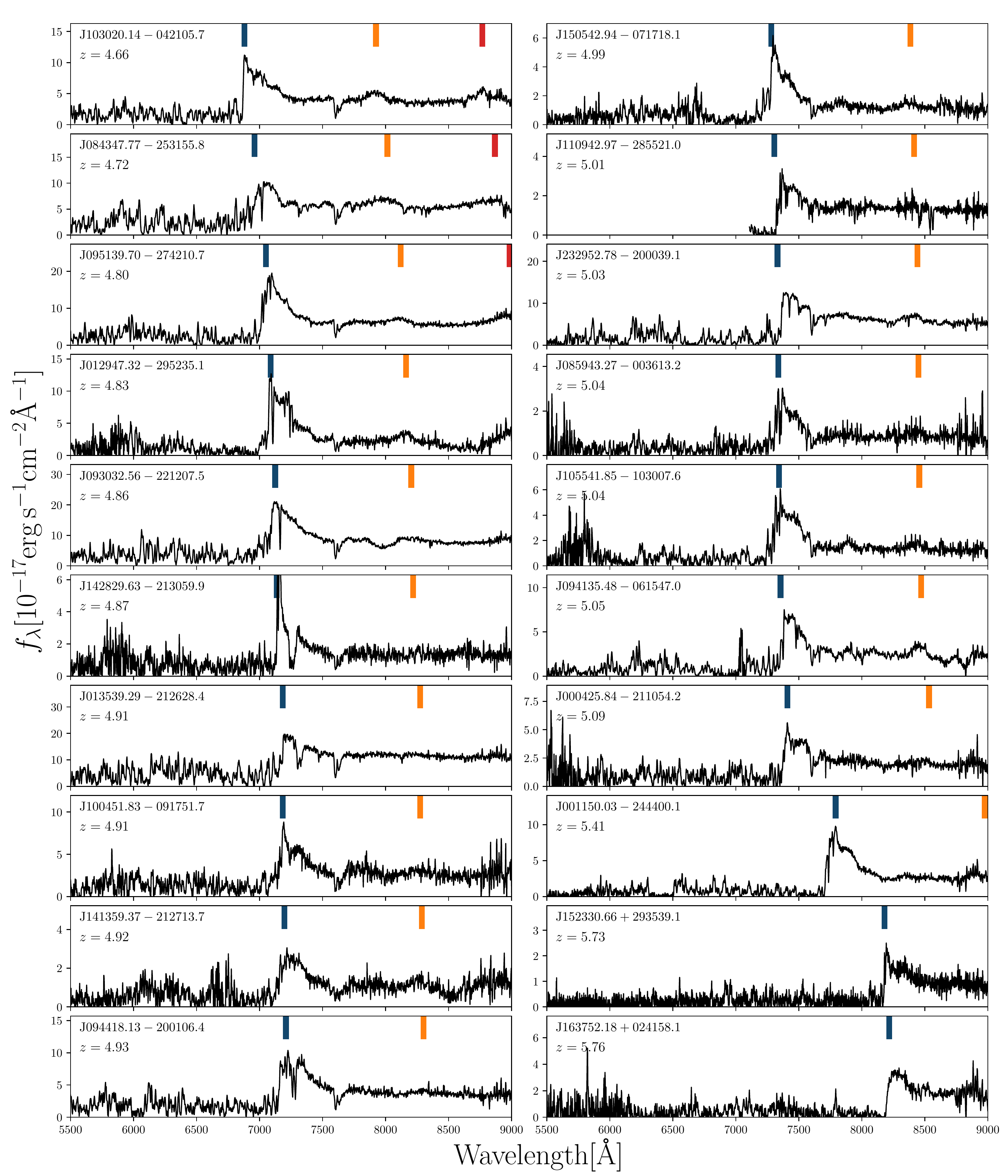}
    \caption{The discovery spectra of the newly identified quasars sorted by spectroscopic redshift. The dark blue, orange and red bars denote the center positions of the broad Ly$\alpha$, Si IV, and C IV emission lines according to the spectroscopic redshift.}
    \label{fig:quasar_spectra}
\end{figure*}

The spectra of the remaining 17 newly discovered quasars at z=4.6-5.1 are also presented in Fig.~\ref{fig:quasar_spectra}. Further information on the individual objects is listed in Table\,\ref{table:observedobjects}. 
These quasars were selected at the low end of our targeted redshift range. One spetrum was obtained with the FORS2 spectrograph on the VLT and all others were obtained with the GoodmanHTS spectrograph on the SOAR telescope. As discussed above not all of them are in our final \textit{high-$z$ candidate set}.

\section{Conclusions} \label{section:conclusion}

The next generation of deep photometric surveys, including the Vera C. Rubin Observatory's Legacy Survey of Space and Time (LSST) and the Euclid Wide Survey, will vastly expand the amount of available data in this field. Quasar selection at $z\approx5-7$ will transition from catalogs of a few hundred objects to large sets that increasingly enable statistical evaluation. This will constrain the statistical properties of quasars and their host galaxies at the time of re-ionization, but it requires robust selection methods that make optimal use of the available data and our evolving understanding of these quasars.

The increase of newly discovered high-redshift quasars ($z>4.7$) over recent years has paved the way to explore high-redshift quasar selection based on supervised machine learning. 

With this work we demonstrate that large enough training samples for quasars and contaminant stars now exist to select and discover high-redshift quasars based on machine learning, in particular, we applied a random forest classification and regression to Pan-STARRS and WISE data. While the need for reasonably sized, spectroscopically confirmed training sets stops this method from finding new quasars at the highest redshift end currently possible, it does show promise to increase the efficiency of the selection up to redshifts of about 6. This can enable the discovery of more quasars per valuable observing time.

Our method also shows promise in finding quasars that would be missed by traditional approaches like color cuts. One of our newly discovered z=5.7 quasars (J152330.66+293539.1) would be rejected by a common cut on the i-z color for $z>5.6$ quasars (see Section~\ref{sec:quasarmissedbycolorcut}). Our \textit{high-$z$ candidate set} contains more promising candidates that would be rejected by that cut. Therefore our random forest approach shows promise to reach higher completeness and is relevant for future quasar luminosity estimates.

Carefully applied supervised machine learning methods to select high-z quasars will be crucial to successfully exploit the combination of future wide-area optical (LSST) and NIR (Euclid) surveys. To fully assess the potential of machine learning quasar selection for LSST and Euclid, applying the same methodology as in this paper to combinations of existing optical and near-infrared surveys (e.g. DES+VHS or KiDS+VIKING) would be an important step, once appropriate training sets are constructed.

In cases where spectroscopic follow-up is no longer viable supervised machine learning methods make it possible to create reliable catalogs of likely quasars. These could be used to put tight constraints on the quasar luminosity function at medium to high redshift in future work.

Nevertheless, our approach presented here has several caveats.
The presented random forest approach does not take into account magnitude errors or make use of the variability information from multi-epoch Pan-STARRS observations. This should be considered in future research. Additionally, the used implementation of random forest can not handle missing values in the data. We work around this by replacing missing values with a lower flux limit. While forced photometry likely would be able to extract additional information, it is beyond the scope of this paper to perform this for all 59 million objects in our Pan-STARRS+WISE catalog data. Our approach also requires the use of large area surveys to ensure enough known quasars are in the survey and can be used to train the random forest. Using simulated quasar photometry the approach could be applied to deeper, but smaller area surveys in future research.
Furthermore, while the efficiency of our test observations is consistent with our estimate, the sample size is quite small. A better confirmation of the novel method to estimate the efficiency could be achieved with more spectroscopic follow-up observations in future work.

We summarize our main conclusions from this work below:

\begin{enumerate}
    \item Using supervised machine learning algorithms like random forests to photometrically select high redshift quasars is a data-driven method that is starting to be competitive with other approaches by making effective use of the rapidly expanding catalogs of spectroscopically confirmed objects.
    \item The main challenges for using random forests or other supervised machine learning approaches are creating a representative training set, getting reliable efficiency estimates and avoiding regions of color space with strong stellar overlap.
    \item We present a new method for estimating the selection efficiency based on the sky distribution of the candidates that can give more realistic estimates, consistent with our test observations.
    \item We showed the effectiveness of our approach through test observations from which we presented 20 new high redshift quasars (17 at $4.6 \leq z \leq 5.5$, 2 at $z\sim 5.7$).
\end{enumerate}

The python code for this project is available under \href{https://github.com/lukaswenzl/High-Redshift-Quasars-with-Random-Forests}{github.com/lukaswenzl/High-Redshift-Quasars-with-Random-Forests}

\acknowledgments


The Pan-STARRS1 Surveys (PS1) and the PS1 public science archive have been made possible through contributions by the Institute for Astronomy, the University of Hawaii, the Pan-STARRS Project Office, the Max-Planck Society and its participating institutes, the Max Planck Institute for Astronomy, Heidelberg and the Max Planck Institute for Extraterrestrial Physics, Garching, The Johns Hopkins University, Durham University, the University of Edinburgh, the Queen's University Belfast, the Harvard-Smithsonian Center for Astrophysics, the Las Cumbres Observatory Global Telescope Network Incorporated, the National Central University of Taiwan, the Space Telescope Science Institute, the National Aeronautics and Space Administration under Grant No. NNX08AR22G issued through the Planetary Science Division of the NASA Science Mission Directorate, the National Science Foundation Grant No. AST-1238877, the University of Maryland, Eotvos Lorand University (ELTE), the Los Alamos National Laboratory, and the Gordon and Betty Moore Foundation.

This publication makes use of data products from the Wide-field Infrared Survey Explorer, which is a joint project of the University of California, Los Angeles, and the Jet Propulsion Laboratory/California Institute of Technology, and NEOWISE, which is a project of the Jet Propulsion Laboratory/California Institute of Technology. WISE and NEOWISE are funded by the National Aeronautics and Space Administration.

Some of the data presented in this paper were obtained from the Mikulski Archive for Space Telescopes (MAST). STScI is operated by the Association of Universities for Research in Astronomy, Inc., under NASA contract NAS5-26555.

This research has made use of the NASA/ IPAC Infrared Science Archive, which is operated by the Jet Propulsion Laboratory, California Institute of Technology, under contract with the National Aeronautics and Space Administration.

Based on observations obtained at the Southern Astrophysical Research (SOAR) telescope, which is a joint project of the Minist\'{e}rio da Ci\^{e}ncia, Tecnologia, Inova\c{c}\~{o}es e Comunica\c{c}\~{o}es (MCTIC) do Brasil, the U.S. National Optical Astronomy Observatory (NOAO), the University of North Carolina at Chapel Hill (UNC), and Michigan State University (MSU).

Funding for SDSS-III and SDSS-IV has been provided by the Alfred P. Sloan Foundation, the Participating Institutions, the National Science Foundation, and the U.S. Department of Energy Office of Science. SDSS-IV acknowledges support and resources from the Center for High-Performance Computing at the University of Utah.  The SDSS-III web site is http://www.sdss3.org/. The SDSS web site is www.sdss.org.

SDSS-III is managed by the Astrophysical Research Consortium for the Participating Institutions of the SDSS-III Collaboration including the University of Arizona, the Brazilian Participation Group, Brookhaven National Laboratory, Carnegie Mellon University, University of Florida, the French Participation Group, the German Participation Group, Harvard University, the Instituto de Astrofisica de Canarias, the Michigan State/Notre Dame/JINA Participation Group, Johns Hopkins University, Lawrence Berkeley National Laboratory, Max Planck Institute for Astrophysics, Max Planck Institute for Extraterrestrial Physics, New Mexico State University, New York University, Ohio State University, Pennsylvania State University, University of Portsmouth, Princeton University, the Spanish Participation Group, University of Tokyo, University of Utah, Vanderbilt University, University of Virginia, University of Washington, and Yale University.

SDSS-IV is managed by the Astrophysical Research Consortium for the 
Participating Institutions of the SDSS Collaboration including the 
Brazilian Participation Group, the Carnegie Institution for Science, 
Carnegie Mellon University, the Chilean Participation Group, the French Participation Group, Harvard-Smithsonian Center for Astrophysics, 
Instituto de Astrof\'isica de Canarias, The Johns Hopkins University, Kavli Institute for the Physics and Mathematics of the Universe (IPMU) / 
University of Tokyo, the Korean Participation Group, Lawrence Berkeley National Laboratory, 
Leibniz Institut f\"ur Astrophysik Potsdam (AIP),  
Max-Planck-Institut f\"ur Astronomie (MPIA Heidelberg), 
Max-Planck-Institut f\"ur Astrophysik (MPA Garching), 
Max-Planck-Institut f\"ur Extraterrestrische Physik (MPE), 
National Astronomical Observatories of China, New Mexico State University, 
New York University, University of Notre Dame, 
Observat\'ario Nacional / MCTI, The Ohio State University, 
Pennsylvania State University, Shanghai Astronomical Observatory, 
United Kingdom Participation Group,
Universidad Nacional Aut\'onoma de M\'exico, University of Arizona, 
University of Colorado Boulder, University of Oxford, University of Portsmouth, 
University of Utah, University of Virginia, University of Washington, University of Wisconsin, 
Vanderbilt University, and Yale University.

This research has made use of the SVO Filter Profile Service (http://svo2.cab.inta-csic.es/theory/fps/) supported from the Spanish MINECO through grant AyA2014-55216

This  work  is  based  on  observations  collected  at  the European  Southern  Observatory  under  ESO  program 105.204A.001.


%

\vspace{5mm}
\facilities{SOAR (GOODMAN), MMT (Red Channel), LBT (MODS), Magellan (FIRE), NOT (NOTCAM), WISE, Pan-STARRS, SDSS}


\software{sklearn \citep{Pedregosa2011}, astropy \citep{2013A&A...558A..33A,2018AJ....156..123A}, python3 \citep{10.5555/1593511}, pandas \citep{mckinney-proc-scipy-2010}, numpy \citep{2011CSE....13b..22V,2020arXiv200610256H}, scipy \citep{scipy2001,2020NatMe..17..261V}, matplotlib \citep{2007CSE.....9...90H}, astroML \citep{2012cidu.conf...47V,2014ascl.soft07018V}, astroquery \citep{2019AJ....157...98G}, sfdmap\footnote{github.com/kbarbary/sfdmap}, IRAF \citep{1986SPIE..627..733T,1993ASPC...52..173T}, MAST\footnote{archive.stsci.edu}, IRSA/ GATOR\footnote{irsa.ipac.caltech.edu/applications/Gator/}, LSD\footnote{github.com/mjuric/lsd}}

\bibliographystyle{aasjournal}
\bibliography{references_ads,references_notinads}







\end{document}